# Electrical control of spins and giant g-factors in ring-like coupled quantum dots

H. Potts[1†*], I–J. Chen[1†], A. Tsintzis[1], M. Nilsson[1], S. Lehmann[1], K. A. Dick[1,2], M. Leijnse[1], C. Thelander[1*]

[1] Division of Solid State Physics and NanoLund, Lund University, SE-221 00 Lund, Sweden

[2] Centre for Analysis and Synthesis, Lund University, SE-221 00 Lund, Sweden

[†] These authors contributed equally to the work. [*] heidi.potts@ftf.lth.se, claes.thelander@ftf.lth.se

**Emerging theoretical concepts for quantum technologies have driven a continuous search for structures where a quantum state, such as spin, can be manipulated efficiently. Central to many concepts is the ability to control a system by electric and magnetic fields, relying on strong spin-orbit interaction and a large *g*-factor. Here, we present a new mechanism for spin and orbital manipulation using small electric and magnetic fields. By hybridizing specific quantum dot states at two points inside InAs nanowires, nearly perfect quantum rings form. Large and highly anisotropic effective *g*-factors are observed, explained by a strong orbital contribution. Importantly, we find that the orbital and spin-orbital contributions can be efficiently quenched by simply detuning the individual quantum dot levels with an electric field. In this way, we demonstrate not only control of the effective *g*-factor from 80 to almost 0 for the same charge state, but also electrostatic change of the ground state spin.**

Quantum dots (QDs) are normally conceived as point-like objects, but can have many different geometries that affect their electronic structure. The development of two-dimensional semiconductor heterostructures provided a particularly flexible material platform, and enabled studies of QDs with various symmetries [1, 2], as well as of molecular states resulting from coupling between two QDs [3, 4]. Another milestone in QD research was the discovery of carbon nanotubes, which offer exceptionally strong confinement, and a special cylindrical geometry with important consequences for spin and orbital interactions [5-7]. Nanowire synthesis further expanded the prospects for controlling orbital and spin states by providing access to low-dimensional narrow bandgap materials, such as InAs and InSb, with inherent strong spin-orbit coupling and confinement effects [8, 9]. These structures have become cornerstones in Majorana research [10-12] and have been considered as a basis for spin qubits.

In this work, we explore the physics of strongly confined QDs coupled in a new geometry, having two connection points instead of just one. We find the unexpected appearance of giant effective *g*-factors ($g^*$) that can be tuned over an exceptionally wide range, which can be explained by formation and quenching of quantum rings. The QDs reside within radial quantum wells in InAs nanowires, close to the nanowire surface [13-15]. Using co-tunneling spectroscopy we explore the energy level structure and its response to a magnetic field when specific orbitals of the two individual QDs align and hybridize to form new states. We find that many interacting orbitals have the expected behavior of strongly coupled QDs, while others behave as quantum rings.

The ring-like states are characterized by vanishing hybridization energies, and have large and highly anisotropic effective g-factors, in very good agreement with recent theoretical predictions for InAs nanowires [16], varying from above 80 in aligned *B*-fields, to below 3 in perpendicular fields. Notably, the electronic structure of the ring-like states is almost identical to that of carbon nanotube QDs, having a nearly fourfold orbital- and spin-degeneracy, broken by spin-orbit interaction [6, 17, 18]. The quality of the rings, quantified by a back-scattering term (δ) connecting states with opposite orbital momentum sign, is similar to values reported only for ultra-clean carbon nanotubes [6].

However, in contrast to carbon nanotubes, we can dramatically affect the electron wave function, and δ, by forming or quenching the ring-like orbitals using electric fields. As a result, we show that $|g^*|$ can be electrostatically controlled from ~80 to ~0 for the same charge state, and the ground state spin can be changed in a constant magnetic field. Previous reports on tuning $g^*$ in QDs cover a much smaller span of modulation [19-21] and often rely on changing the charge state [8, 9].

We develop theoretical models showing excellent agreement with the experimental data. Using perturbation theory, we furthermore find that the contribution from the inter-dot tunnel-coupling to the hybridization gap can effectively cancel if an odd orbital on one QD hybridizes with an even orbital on the other QD. This method of creating high quality quantum rings is thus generic, and opens up for $g^*$ manipulation and spin control in many material systems.

**Coupling strongly confined quantum dots**

The quantum rings studied in this work appear inside quantum wells (QWs) formed during epitaxial growth of InAs nanowires [22]. The nanowire crystal phase is controlled to form structures similar to that shown in Fig. 1a, having a 5 nm long zinc-blende (ZB) segment sandwiched between two wurtzite (WZ) segments of about 30 nm. Due to a conduction-band offset, the WZ segments confine electrons in the centre ZB segment, resulting in a thin QW accessed by two tunnel barriers. We study the properties of the QW by first fabricating source-, drain- and gate-electrodes, as indicated in Fig. 1b, followed by transport measurements in a dilution refrigerator equipped with a vector magnet.

The electrostatics in the QW is manipulated through voltages applied to two side-gates ($V_L$ and $V_R$) and a global back-gate ($V_{BG}$), such that two QDs form, which are parallel-coupled to source and drain (Fig. 1c). The existence of a double QD is evidenced by the characteristic honeycomb diagram shown in Fig. 1d. Here, the strong confinement allows extraction of the electron population on each QD, and provides a clear spin-pairing of orbital states, such that the orbital number ($O_L$, $O_R$) can be extracted. For this system, QD occupancies and inter-dot tunnel coupling are tunable over a wide range owing to the relatively rigid crystal phase confinement [13].

Previous works on these structures focused on the first orbital crossing ($O_{L,R} = 1$), for which each QD contains zero to two electrons [13-15]. In this work, we instead investigate interactions of higher orbitals. The crossing of ($O_L$, $O_R$) = (2, 3) (highlighted in Fig. 1d) is treated in the main text, whereas results from other crossings (both from the same and another device) are shown in the Supplementary Information (SI). The electron population on each QD for this crossing is indicated in Fig. 1e. However, since filled orbitals are considered not to interact with other electrons, we will instead refer to the one- two- and three-electron regime (1e, 2e, 3e) in the rest of the article. We note that the conductance lines outlining the (2,3) crossing show very sharp corners. If interpreted in a standard double-QD picture, this would imply the absence of inter-dot tunnel coupling, which is in stark contrast to the clear hybridization gap (rounded corners) of the neighboring (2,2) crossing.

Transport is investigated by recording differential conductance, $dI/dV_{ds}$, as a function of source-drain voltage ($V_{ds}$) and either $V_{L,R}$ or magnetic field ($B$). The $V_{L,R}$ vectors are represented as green, red and yellow vectors in Fig. 1e, while the direction of the $B$-field with respect to the nanowire is defined in Fig 1b. A measurement recorded along the green gate vector is shown in Fig. 1f for $B_{||} = 0.05$ T aligned with the nanowire. Sequential tunneling here provides weakly outlined Coulomb diamonds representing the 1e, 2e, and 3e regime. Within the Coulomb diamonds, the electron number in the QD system is constant, and the onset of co-tunneling processes allows extraction of excited state energies with respect to the ground state. In the 1e and 3e regime, the 1$^{st}$ excited state corresponds to a $B$-field induced splitting of the ground state. Effective $|g^*|$-factors of approximately 64 and 40 can be extracted from $\delta E = |g^*|\mu B$ for the 1e and 3e regime, respectively. These values are significantly larger than for bulk InAs (-14.9), which indicates a very strong orbital contribution [8, 16].

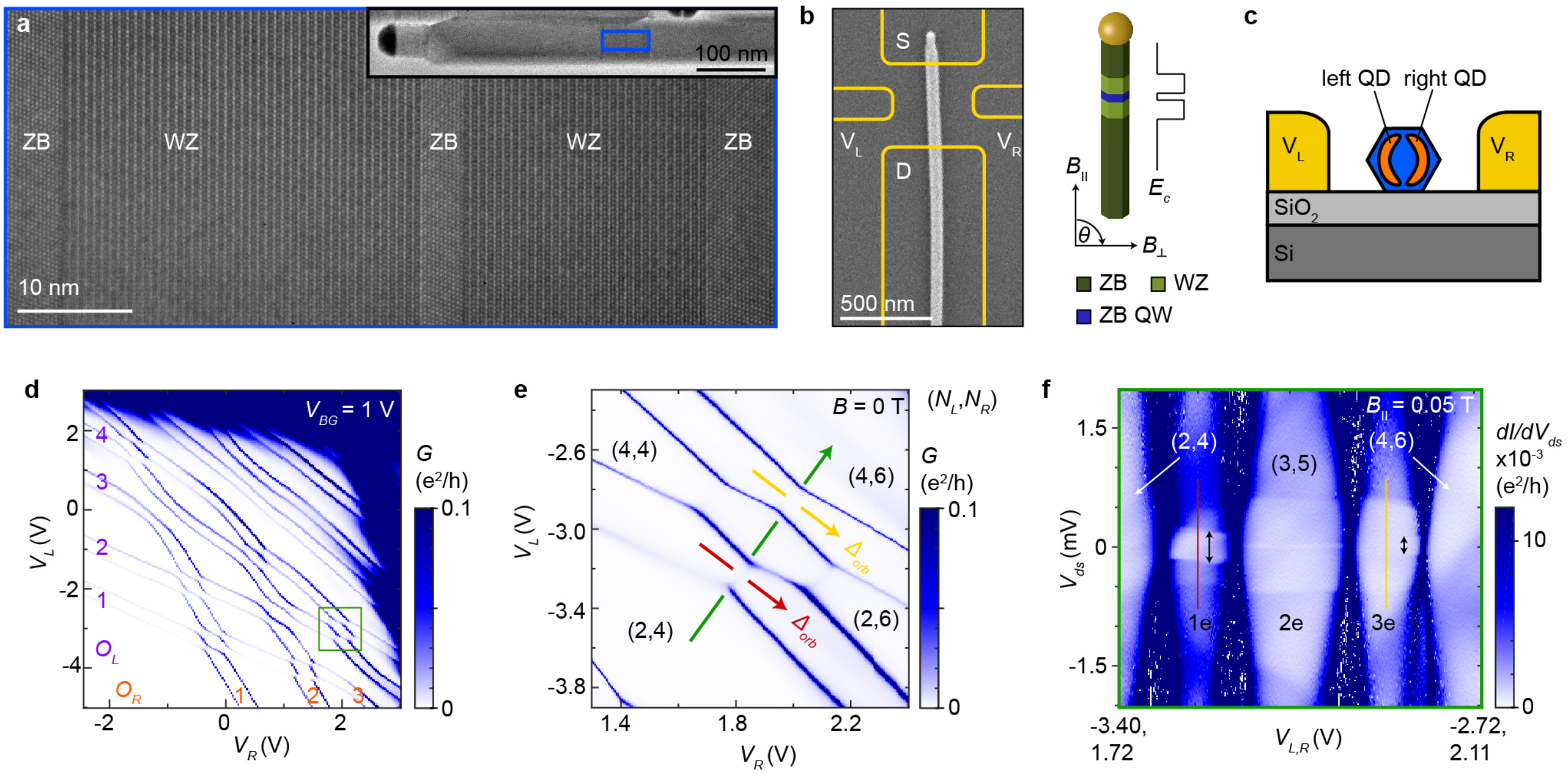

*Figure 1* ***Formation of parallel-coupled QDs based on InAs nanowires. a*** *Transmission electron micrograph of a representative nanowire. The centre zinc blende (ZB) segment acts as a quantum well accessed by wurtzite (WZ) tunnel barriers.* ***b*** *Left: SEM of the studied nanowire, overlaid with the contact design. Right: Schematic of the crystal structure and resulting conduction-band alignment.* ***c*** *Side-view illustration of the formation of two parallel-coupled QDs.* ***d*** *Conductance G as a function of side-gate voltages ($V_L$, $V_R$), where the orbital numbers ($O_L$,$O_R$) are indicated.* ***e*** *Magnification of the ($O_L$,$O_R$)=(2,3) crossing. Green, red, and yellow arrows indicate important gate vectors, and ($N_L$, $N_R$) represents the electron population on the left and right QD.* ***f*** *Measurement of $dI/dV_{ds}$ versus $V_{ds}$ along the green gate vector for $B_{||}$ = 0.05 T. The Zeeman splitting of the ground state $2\delta E$ is indicated with arrows in the 1e and 3e regime.*

## Forming a quantum ring

To understand the orbital contribution to $g^*$ we first study the evolution of states as functions of $B$-field strength and direction. We start by investigating the 1e regime at a $V_{L,R}$ where $O_L = 2$ and $O_R = 3$ are degenerate (referred to as zero detuning, $\Delta_{orb} = 0$). Figure 2a shows transport as a function of $V_{ds}$ and $B_{||}$. Three excited states evolving with different slopes can be identified, indicating that the underlying process is not just a standard Zeeman splitting of spin states.

In order to understand this behaviour, we implemented a three-dimensional simulation of the device structure, including a Rashba-type spin-orbit interaction term. By solving the single-electron

Schrödinger equation, we can extract the energies of the ground state (GS) and the excited states (ESs). Figure 2b shows the states in the 1e regime at the crossing of the 2$^{nd}$ and 3$^{rd}$ orbitals. Hybridization of these orbitals leads to four new states, two of which increase rapidly in energy with $B_{||}$, and two of which decrease. This can be understood in the following way: The hybridized wave function resembles a ring in which electrons can orbit clockwise or anti-clockwise (green/red), similar to what is known from carbon nanotube QDs [6, 17]. Figure 2c shows calculated onsets of inelastic co-tunneling processes through excited states, which can be directly compared to our experimental results. Excellent quantitative agreement was obtained by adjusting the nanowire surface charge, the effective spin $g$-factor ($g^*_{spin}$), and the SOI strength, while keeping these within the ranges of literature values. A nearly circular symmetry can also be identified in the calculated probability density distribution of the lowest energy state as shown in Fig. 2d. We obtain qualitatively similar results using a simple tight-binding model of a ring broken by two barriers (details of the models are provided in SI).

A finite $B_{||}$ couples to both the spin and the orbital momentum of the ring-like states, where $g^*$ is given by $g^* \approx g^*_{spin} \pm g^*_{\mathrm{orbit}}$. Since the effective orbital $g$-factor ($g^*_{\mathrm{orbit}}$) dominates, the two clockwise states decrease in energy with $B_{||}$, whereas the anti-clockwise states increase. For a perfect ring, the four states at $B = 0$ are split by the spin-orbit interaction energy ($\Delta_{SOI}$) into two pairs, similar to the Kramers doublets (K↓, K'↑) and (K↑, K'↓) for carbon nanotubes. We now define $g_1$* and $g_2$* as the effective $g$-factor of the lower and higher Kramers doublet, respectively. A positive (negative) value for g* indicates that the lower-energy state is spin-down (spin-up). Experimental values of $g_1^* \approx 59$ and $g_2^* \approx -83$ are extracted from Fig. 2a, which roughly corresponds to $g^*_{spin} = -12$ and $g^*_{orbit} = 71$. We note that signs of $g_1$* and $g_2$* depend on the sign of SOI. In our case $g_1$* is positive as the (positive) orbital contribution is added to the (negative) spin-contribution, while $g_2$* is negative and larger in magnitude because both contributions have negative sign.

In a ring with disorder, states with different orbital momentum sign are coupled, resulting in an energy split (δ) related to backscattering. The energy gap at $B = 0$ is then given by $\Delta E_{B=0} = \sqrt{\Delta_{SOI}{}^2 + \delta^2}$. From the high resolution experimental data in Fig. 2e, we observe $\Delta E_{B=0} \approx 240$ μeV, and $\delta \approx 50$ μeV can be extracted from the anti-crossing at $B_{||} \approx 55$ mT. This allows to estimate $\Delta_{SOI} = 235$ μeV, showing that $\Delta E_{B=0}$ is dominated by SOI, which confirms the high ring quality. Interestingly, carbon nanotubes with a corresponding parameter $\delta_{KK'} \approx 65$ μeV have been referred to as 'ultraclean' [6]. Going to even higher $B_{||}$, we ultimately find a change of the spin ground state at $B_{||} \approx 0.35$ T (Fig. 2a), where the Zeeman splitting of clockwise states (green) overcomes the SOI-induced energy gap.

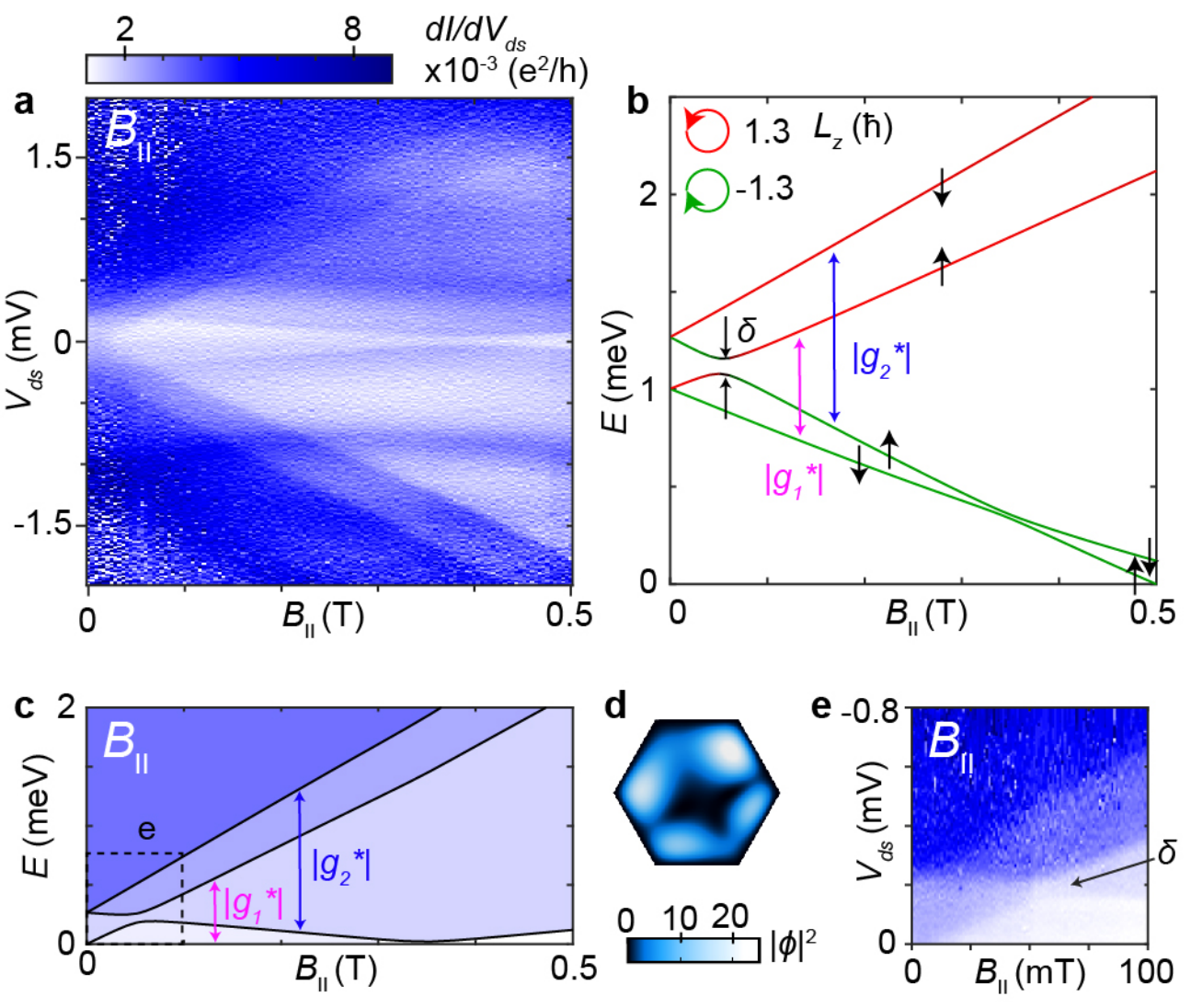

*Figure 2* ***Spectroscopy of the 1e regime at zero detuning and*** $\boldsymbol{B_{||}}$**.** ***a*** *Measurement of* $dI/dV_{ds}$ *versus* $V_{ds}$ *and* $B_{||}$*.* ***b*** *Numerical calculation of state energies as a function of* $B_{||}$. The colour represents the calculated orbital angular momentum $L_z$, and $|g_1^*|$ *and* $|g_2^*|$ *correspond to the B-field induced splitting of the two Kramers pairs. The spin arrows indicate the alignment of the spin with respect to the nanowire axis (up/down corresponds to parallel/anti-parallel). Due to the presence of SOI, the arrows only indicate the approximate spin direction.* ***c*** *Calculated onsets of co-tunneling processes starting from the lowest energy state as a function of* $B_{||}$ *(artificial colouring).* ***d*** *Absolute square of the wave function of the lowest energy state at infinitesimal* $B_{||}$*-field and* $\Delta_{orb} = 0$*.* ***e*** *Same as panel a, zooming in on the small anti-crossing (*$\delta$*) due to coupling of states with same spin but different orbital momentum sign.*

Next, we investigate the evolution of states with $B_\perp$ applied perpendicular to the nanowire axis. With increasing $B_\perp$, a gap slowly opens between the ground state and the first excited state (ES1), as shown in Fig. 3a,b. The weak $B_\perp$ dependence is a result of the mixing of spin and orbital states due to SOI, since the avoided crossing strongly suppresses Zeeman splitting at small $B_\perp$ [6]. Theory predicts that the GS-ES1 splitting converges to $\delta$ when both states have the same spin orientation at higher $B_\perp$. This strong non-linearity makes it difficult to define $g^*$ of the ring-like state for $B_\perp$.

The exceptionally strong dependence of the state energies on the $B$-field direction ($\theta$) can be visualized by rotating $B$ in the plane of the nanowire. An excellent agreement between experimental and corresponding simulation data is shown in Figs. 3c,d for $B = 0.2$ T. From the experiment we extract $|g_1^*|$ and $|g_2^*|$ as a function of $B$-field direction, plotted in Fig. 3e.

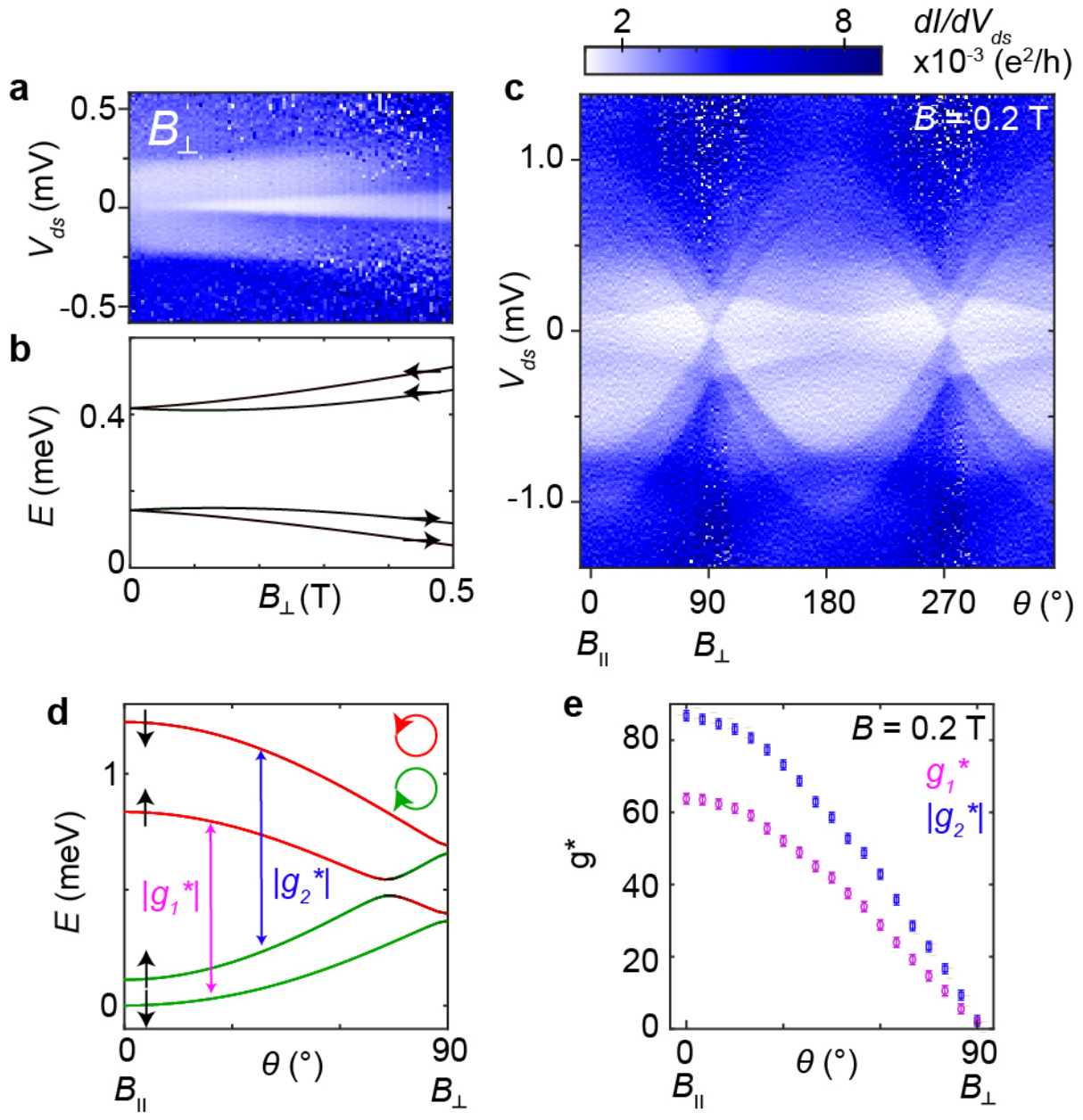

*Figure 3* ***Spectroscopy of the 1e regime at zero detuning for different B-field directions. a*** *Measurement of $dI/dV_{ds}$ versus $V_{ds}$ and $B_\perp$.* ***b*** *Numerical calculation of the state energies as functions of $B_\perp$. The spin arrows indicate the alignment of the spin with respect to the external B-field.* ***c-d*** *Measurement of $dI/dV_{ds}$ and numerical calculation as a function of B-field direction for 0.2 T. The strong orientation dependence is a result of the orbital contribution to g*.* ***e*** *Experimental values of $|g_1^*|$ and $|g_2^*|$ as functions of B-field direction.*

## Quenching the ring

So far we have discussed the properties of high-quality quantum rings when two orbitals of different QDs align ($\Delta_{orb} = 0$). In the following, we investigate detuning of the involved orbitals with an electric field. This allows us to control the ring quality *in situ*, which has not been demonstrated for carbon nanotubes.

In Fig. 4a we present transport as a function of level detuning in the 1e regime (red vector in Fig. 1e) for $B_{||}$ = 0.1 T. Detuning is here defined as the change in side-gate voltages with respect to orbital degeneracy ($V_{R0}$, $V_{L0}$), and can be converted to an energy $\Delta_{orb}$ (see SI). Comparing with simulations (Fig. 4b), we see that the splitting between the ground state and the first excited state strongly depends on detuning and rapidly decays when going away from orbital degeneracy. This indicates that the ring state quenches with detuning.

The ability to control the magnetic field response ($g^*$) using an electric field is important in many proposed concepts for spin manipulation, such as in spintronics and $g$-tensor modulation techniques [20, 23, 24]. Based on a linear approximation, we have extracted $|g^*|$ from the GS-ES1 transition for $B_{||} = 0.1$ T, and $B_{||} = 0.04$ T (Fig. 4a and SI). The results in Fig. 4c show that $|g^*|$ can be electrostatically tuned from above 80 down to almost 0, with a peak $dg^*/d\Delta V_{L,R} > 1000$ $V^{-1}$, which is very attractive for *g*-tensor modulation.

In Fig. 4d we present a larger detuning range for a measurement at $B_{||} = 0.2$ T. By gradually destroying the quantum ring, and thereby decreasing its orbital contribution, the GS-ES1 energy gap approaches zero near $\Delta V_{R,L}$= -0.2 V, followed by a ground state change. This can be understood by considering that the ground state at orbital degeneracy is spin down (for $B_{||} < 0.35$ T), whereas the unperturbed orbital $O_L$ has a spin-up GS. Theory predicts that the ground-state change should show an exact crossing if the external *B*-field is aligned with the spin-orbit field (in this case spin is a good quantum number, regardless of disorder). A similar 1e spin ground state change was demonstrated by Hauptmann *et al.* using electrostatic manipulation of the exchange field from ferromagnetic contacts to a carbon nanotube QD [25].

Next, we investigate the effect of detuning for the case of $B_{\perp}$. Figures 4e-f show the experimental data and corresponding simulation for $B_{\perp} = 0.1$ T. As previously discussed, mixing of spin and orbital states due to SOI leads to a suppression of the Zeeman splitting at orbital degeneracy. Near orbital degeneracy, a dip in the GS-ES1 energy gap is therefore observed both in the experimental and simulation data. Extracted values for $|g^*|$ as functions of detuning are presented in Fig. 4g. The extracted $|g^*| \sim 3$ at orbital degeneracy should be considered an upper bound as it is difficult to find the specific angle where no external field penetrates the ring. When the ring is quenched by detuning, $|g^*|$ approaches 7.5, which is in line with typical values observed in InAs QDs [14, 26]. The peak of $dg^*/d\Delta V_{L,R}$ corresponds to >100 $V^{-1}$.

The calculated probability density of the lowest energy state for three different detuning values (Fig 4h) shows a ring-like wave function at $\Delta_{orb} = 0$, whereas $\Delta_{orb} \neq 0$ shifts the wave function to either side and clearly destroys the ring symmetry.

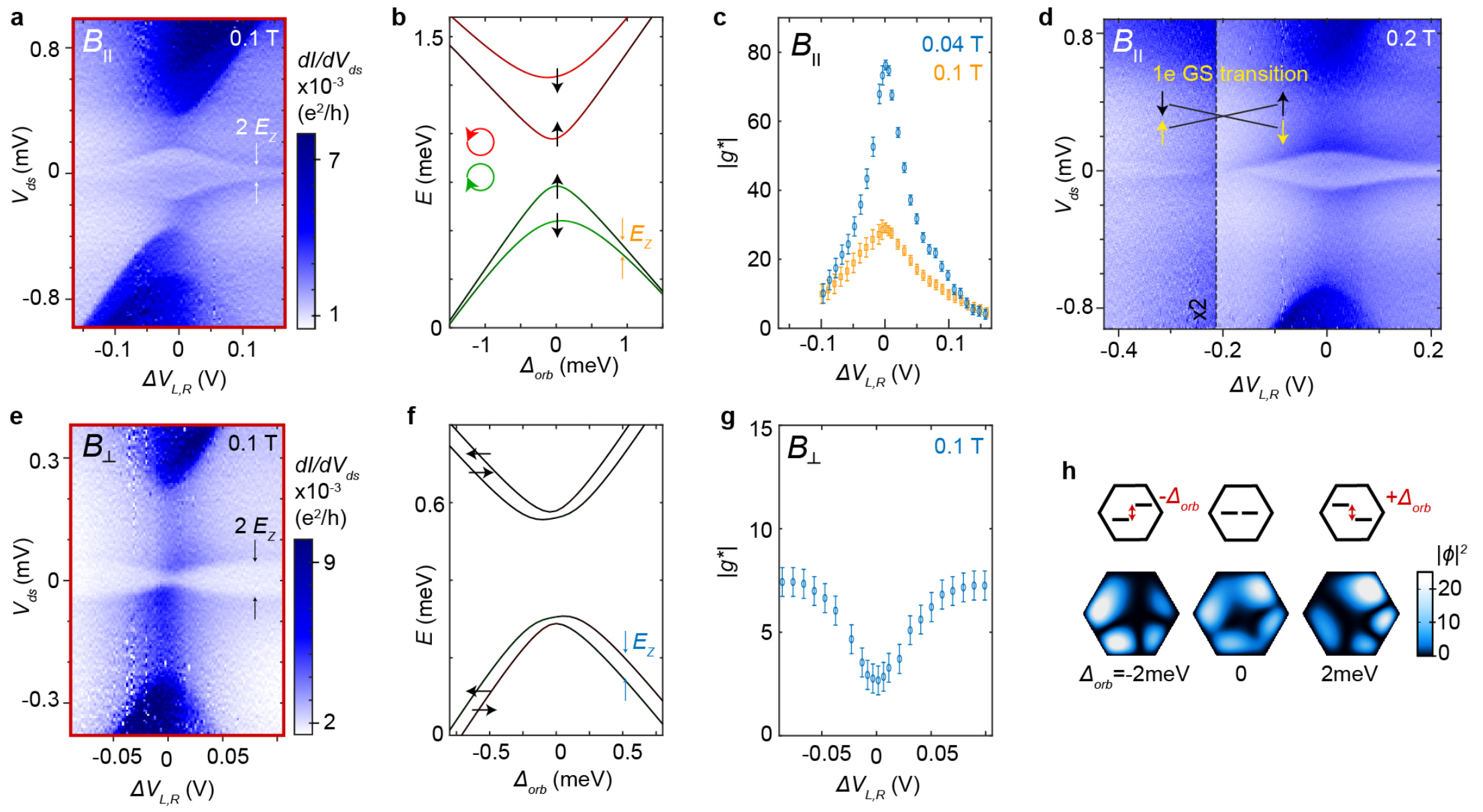

*Figure 4* ***Quenching the 1e quantum ring by detuning left and right QD orbitals. a*** *Measurement of $dI/dV_{ds}$ versus $V_{ds}$ recorded along the red detuning vector (c.f. Fig. 1e) for $B_{\|} = 0.1$ T. Detuning is defined as $\Delta V_{L,R}=(V_R-V_{R0})-(V_L-V_{L0})$, where $(V_{R0},V_{L0})$ are the side-gate voltages when the left and right QD levels are degenerate.* ***b*** *Numerical calculation of the states for $B_{\|} = 0.1$ T. Here detuning refers to the extrapolated energy difference between the unperturbed orbitals in the left and right QD (see schematic in Fig. 4h).* ***c*** *Experimental values of $|g^*|$ for $B_{\|} = 0.04$ and 0.1 T, extracted from the GS-ES1 transition using on a linear approximation. In the case of $B_{\|} = 0.04$ T, $|g^*|$ at zero detuning corresponds to $|g_1^*|$, while we find a strongly reduced value at $B_{\|} = 0.1$ T due to the orbital change of the 1$^{st}$ excited state.* ***d*** *Same as panel* ***a****, but for $B_{\|} = 0.2$ T and a larger detuning range, showing a detuning-induced spin-change of the 1e ground state.* ***e-f*** *Measurement of $dI/dV_{ds}$ and numerical calculation as a function of detuning for $B_{\perp} = 0.1$ T, where SOI from the ring supresses Zeeman splitting.* ***g*** *Experimental values of $g^*$ for the GS-ES1 transition at $B_{\perp} = 0.1$ T.* ***h*** *Calculated probability density of the lowest energy state at infinitesimal $B_{\|}$-field for three different detuning energies. The error bars in c and g correspond to the uncertainty in measurement resolution and data extraction.*

## 3-electron regime

In the following, we discuss the 3e regime, which typically is equivalent to the 1e regime in spin-degenerate two-orbital systems due to particle-hole symmetry. However, this is not the case in ring-like QDs with strong SOI [6]. The specific ordering of the Kramers pairs is here determined by the signs of $\Delta_{SOI}$ and $g^*_{spin}$. The orbital crossing (2,3) has a smaller splitting of the lower pair compared to the upper pair, which is consistent with $\Delta_{SOI} > 0$ and $g^*_{spin} < 0$. This, however, is different for other orbital crossings, as shown in SI, where also $\Delta_{SOI} < 0$ is found.

The 3e case can be understood by considering that the first three states are filled, and transitions to the 4$^{th}$ state are probed. Opposite to the 1e case, the GS-ES1 energy gap at orbital degeneracy should therefore continuously increase with $B_{\parallel}$. Spectroscopic data recorded as a function of $B_{\parallel}$ (Fig 5a) and $B$-field angle (Fig. 5b) agrees with this interpretation. Figure 5c shows the characteristic increase of the ES1-GS energy upon formation of the ring-like state. More experimental data for both $B_{\parallel}$ and $B_{\perp}$ can be found in SI. Figure 5d shows $|g^*|$ extracted as a function of detuning for the 3e regime. Similar to the 1e case, $|g^*|$ recorded for $B_{\parallel} < 55$ mT can directly be interpreted as $|g_2^*|$. The maximum value of $|g_2^*| \approx 83$ at orbital degeneracy matches with the estimation based on the $B_{\parallel}$-sweep of the 1e case. However, as the 4th state is already spin-down at orbital degeneracy, no change in ground state spin is observed with detuning, which also reduces the peak $dg^*/d\Delta V_{L,R} > 800$ V$^{-1}$ compared to the 1e case.

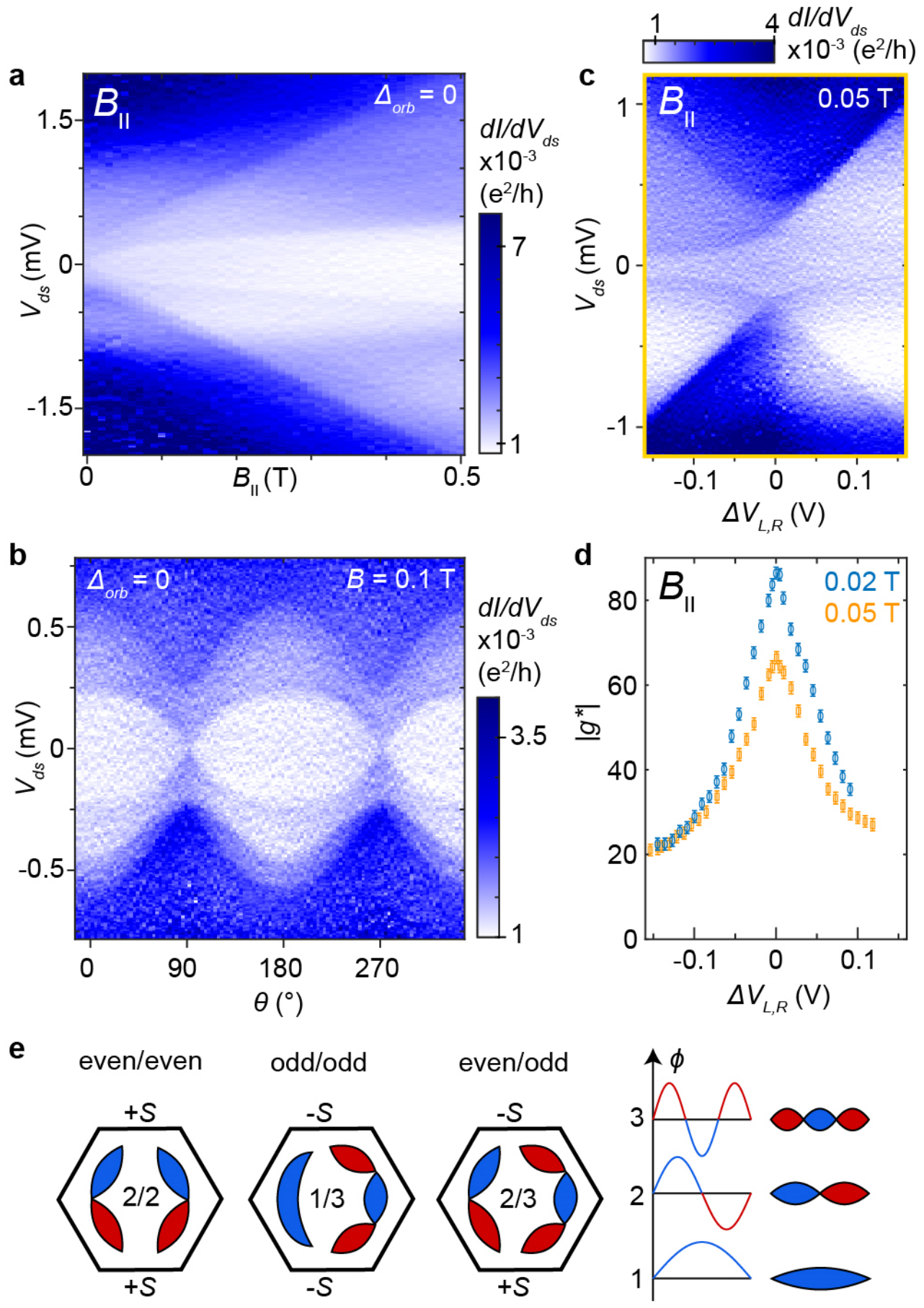

*Figure 5* ***3e regime and orbital requirements for ring formation.*** ***a*** *Measurement of dI/dV$_{ds}$ versus $V_{ds}$ and $B_{\parallel}$ at zero detuning. The SOI breaks e-h symmetry, leading to an increase of all three excited states with $B_{\parallel}$ in the 3e regime.* ***b*** *Measurement of dI/dV$_{ds}$ as a function of B-field direction for B = 0.1 T at $\Delta_{orb}$ = 0.* ***c*** *Measurement of dI/dV$_{ds}$ when detuning the orbitals along the yellow gate vector (c.f. Fig. 1e) for $B_{\parallel}$ = 0.05 T.* ***d*** *Experimental values of $|g^*|$ for the GS-ES1 transition based on a linear approximation. In the case of $B_{\parallel}$ = 0.02 T, the extracted $|g^*|$ corresponds to $|g_2^*|$, while we find a reduced value at $B_{\parallel}$ = 0.05 T due to the avoided orbital crossing near this $B_{\parallel}$-field.* ***e*** *Schematic representation of the wave functions in crossings of even/even, odd/odd, and even/odd orbitals. Nearly perfect rings can only be observed in the case of even/odd combinations where the overlap integral (S) has different sign at the two connection points.*

Finally, we address the general question of the requirements necessary to observe these ring-like states. Based on our tight-binding model and a simpler DQD model (details in SI) we find that a nearly perfect ring can form even in the presence of significant tunnel barriers between the QDs if two conditions are fulfilled: 1) the tunnel coupling strength at the two connection points of the QDs are identical, and 2) an even orbital in one QD is energetically aligned with an odd orbital in the other QD. An even/even or odd/odd combination of orbitals (first two panels in Fig. 5e) leads to a large energy gap between the lower and the higher Kramers pair, similar to standard bonding and anti-bonding orbitals of QDs which are connected in one point. However, in the case of an even/odd combination (third panel of Fig. 5e), the hybridization gap vanishes because of the different signs of the overlap integrals ($S$) at the two QD connection points. The resulting four degenerate states now split in the presence of SOI into two Kramers pairs, each of which show ring-like behaviour. There is thus no strict requirement for a particular material, and other coupled QD systems should display similar physics. This may extend the possibilities for spin manipulation and tuning of g* in QD systems with longer spin coherence times, and allow for strong coupling of spins to cavity photons.

**Methods**

InAs nanowires with controlled crystal phase structure were grown by metal-organic chemical vapour deposition (MOCVD) and analysed in a Hitachi HF3300S transmission electron microscope operated at 300 keV. Select nanowires were deposited with a micromanipulator onto highly doped silicon samples with 200 nm $SiO_2$. Contacts were then fabricated using e-beam lithography followed by native oxide removal in HCL(37%):$H_2O$ (1:20) for 15 s and evaporation of 20/80nm Ni/Au. More details on the sample fabrication can be found in Ref. 13. All experimental data is obtained from asymmetrically biased DC measurements performed in a dilution refrigerator at an electron temperature <100 mK. The differential conductance was calculated using the gradient function in Matlab. In some cases, noise from the raw data was reduced by a moving average (smooth function) or a local regression (rlowess function). The values of $g^*$ are determined based on a linear approximation. The $B_{||}$-field dependence (Fig. 2a) allows an accurate extraction of $g_1^*$ and $g_2^*$, by taking data points which are far away from the hybridization gaps. In the case of the detuning dependence (Figs. 4c,g and 5d), the reported $g^*$ corresponds to the energy gap between the ground state and first excited state, and is therefore influenced by the hybridization of states.

## Acknowledgements

The authors thank A. Burke for experimental support. Financial support was received from the Knut and Alice Wallenberg Foundation (KAW), the Swedish Research Council (VR), the Crafoord Foundation and NanoLund. H.P. acknowledges the Swiss National Science Foundation (SNSF) via Early PostDoc Mobility.

## Author contributions

H.P., I-J.C. and C.T. did the transport measurements, analysis and figure preparation. I-J.C., A.T. and M.L. developed the theoretical models. H.P. and M.N. fabricated the samples. S.L. and K.A.D. did the epitaxy and structural characterization. All authors contributed to discussions and writing of the manuscript.

## Additional information

The authors declare no competing financial interest. Correspondence and requests for materials should be addressed to H.P. or C.T.

# Supplementary Information

# Electrical control of spins and giant g-factors in ring-like coupled quantum dots

H. Potts[1†*], I–J. Chen[1†], A. Tsintzis[1], M. Nilsson[1], S. Lehmann[1], K. A. Dick[1,2], M. Leijnse[1], C. Thelander[1*]

[1] Division of Solid State Physics and NanoLund, Lund University, SE-221 00 Lund, Sweden

[2] Centre for Analysis and Synthesis, Lund University, SE-221 00 Lund, Sweden

[†] These authors contributed equally to the work. [*] heidi.potts@ftf.lth.se, claes.thelander@ftf-lth.se

## Contents

## Three-dimensional numerical simulation of single-electron states

In this section, we discuss the theoretical model used in the numerical simulation of single-electron states under different electrostatic detuning and magnetic field. Under the effective mass approximation, the Hamiltonian of a single electron in a magnetic field can be expressed as[1]

$$H = \frac{(-i\hbar\nabla + e\vec{A})^2}{2m^*} + V(r) + \frac{g^*_{spin}}{2}\mu_B \vec{\sigma}\cdot\vec{B} + H_{SO}, \qquad \text{(S1)}$$

with electron effective mass $m^*$, elementary charge $e$, electron spin $g$-factor $g^*_{spin}$, Bohr magneton $\mu_B$, Pauli vector $\vec{\sigma}$, magnetic vector potential $\vec{A} = (A_x, A_y, A_z)$, and magnetic field $\vec{B} = \nabla \times \vec{A}$. $V(r)$ and $H_{SO}$ are the electrostatic potential and the spin-orbit (SO) interaction term, respectively. $V(r)$ can be obtained by solving the Poisson equation

$$\nabla \cdot \left(\varepsilon_r \varepsilon_0 \nabla V(r)\right) = \rho, \tag{S2}$$

with dielectric constant $\varepsilon_r$, vacuum permittivity $\varepsilon_0$, and total volume charge density $\rho$.

Here we consider Rashba-type spin-orbit coupling: electric field in the direction $\vec{e}$ results in SO coupling of the form[1,2]

$$H_{SO,R} = \frac{\alpha}{\hbar}(\vec{e} \times \vec{p}) \cdot \vec{\sigma}, \tag{S3}$$

where $\alpha$ is the Rashba parameter and $\vec{p}$ is the electron kinetical momentum. For the simulation results presented in the article, we assume the electric field to be along the nanowire axis ($\hat{z}$). Therefore, we can explicitly express the SO coupling term as

$$H_{SO} = \alpha(\sigma_x \left(-i\frac{\partial}{\partial y} + \frac{eA_y}{\hbar}\right) + \sigma_y \left(-i\frac{\partial}{\partial x} + \frac{eA_x}{\hbar}\right)). \tag{S4}$$

It is however worth noting that a quantitatively similar result was achieved with electric fields in the nanowire radial direction (not shown).

We solve the differential equations (Eq. S1 and S2) based on the finite-element method in COMSOL. First, we obtain $V(r)$ by solving Eq. S2 for the device structure shown in Fig. S1. Afterwards, $V(r)$ is used as an input and Eq. S1 is solved for the InAs nanowire structure colored in blue and green in Fig. S1. Here the Poisson equation and the single electron Hamiltonian are not solved self-consistently, therefore electron-electron interaction is neglected and the calculated spectrum corresponds to the single-electron energy states of an empty quantum dot.

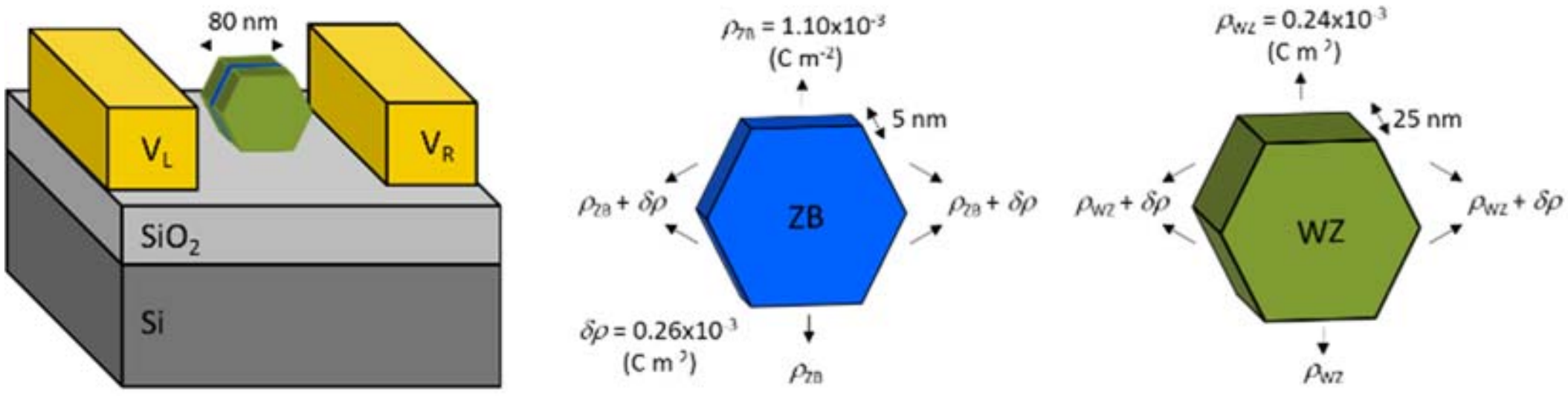

*Figure S1 Device structure used in the numerical simulation. The surface charge on the side and top/bottom facets in zinc blende ZB (blue) and wurtzite WZ (green) InAs are indicated. A lower surface charge is set for wurtzite as found in Ref. 3. An extra positive surface charge density of $\delta\rho = 0.26\text{x}10^{-3}$ C m$^{-2}$ is added to all the side facets in the simulation to fit the experimental data.*

In the simulation an extra positive surface charge density $\delta\rho$ is assigned to the side facets, which leads to electrostatic potential barriers near the top and bottom facets and parallel-coupled double quantum dot states. Several parameters are tuned to match the state energies of the (2,3) crossing (at zero detuning, $\Delta_{orb} = 0$) and their evolution with external magnetic field along the nanowire axis direction $B_{||}$. More specifically, $\delta\rho$ and the gate configurations (set to approximately the experimental values) determine the circular symmetry of $V(r)$ in the nanowire and thus the backscattering of the orbiting electrons, which is characterized by the energy split $\delta$ (indicated by pink arrows in Fig. S2(b)). In addition, they also determine the orbital angular momentum $L_z$ of the states. The angular motion of the electron couples to the external magnetic field and contributes to the $g$-factor with

$$g^*_{orbital} = \frac{\mu_B}{m^*\hbar} L_z, \qquad \text{(S5)}$$

which increases with decreasing electron effective mass $m^*$. In the simulation, by adjusting the surface charges ($\rho_{ZB}$ and $\rho_{WZ}$) and the zinc blende-wurtzite conduction band offset, we can modify the relative probability density distribution of the electronic states in the zinc blende ($m^* = 0.026$ $m_e$[5]) and wurtzite ($m^* = 0.037$ $m_e$[6]) segments to match the simulated $g_{orbital}^*$ (at given $\delta$ and $L_z$) with the experimental values. The effective $g$-factor ($g^*$) is approximately given by $g^* \approx g^*_{spin} \pm g^*_{orbit}$, and therefore $g^*_{spin} \approx (g_1^* + g_2^*)/2$. Finally, $\alpha$ is adjusted to match the energy gap at $B_{||} = 0$ ($\Delta E_{B=0}$, indicated by blue arrows in Fig. S2(b))) and the value of $B_{||}$ when states with opposite signs of $L_z$ anti-cross (indicated by pink arrows). The calculated values of $\delta$, $\Delta E_{B=0}$, $g_1^*$, $g_2^*$, and the relevant material parameters are shown in Table S1. The same parameters are used to simulate the evolution of the state energies at different detuning and external magnetic field as shown in the main text.

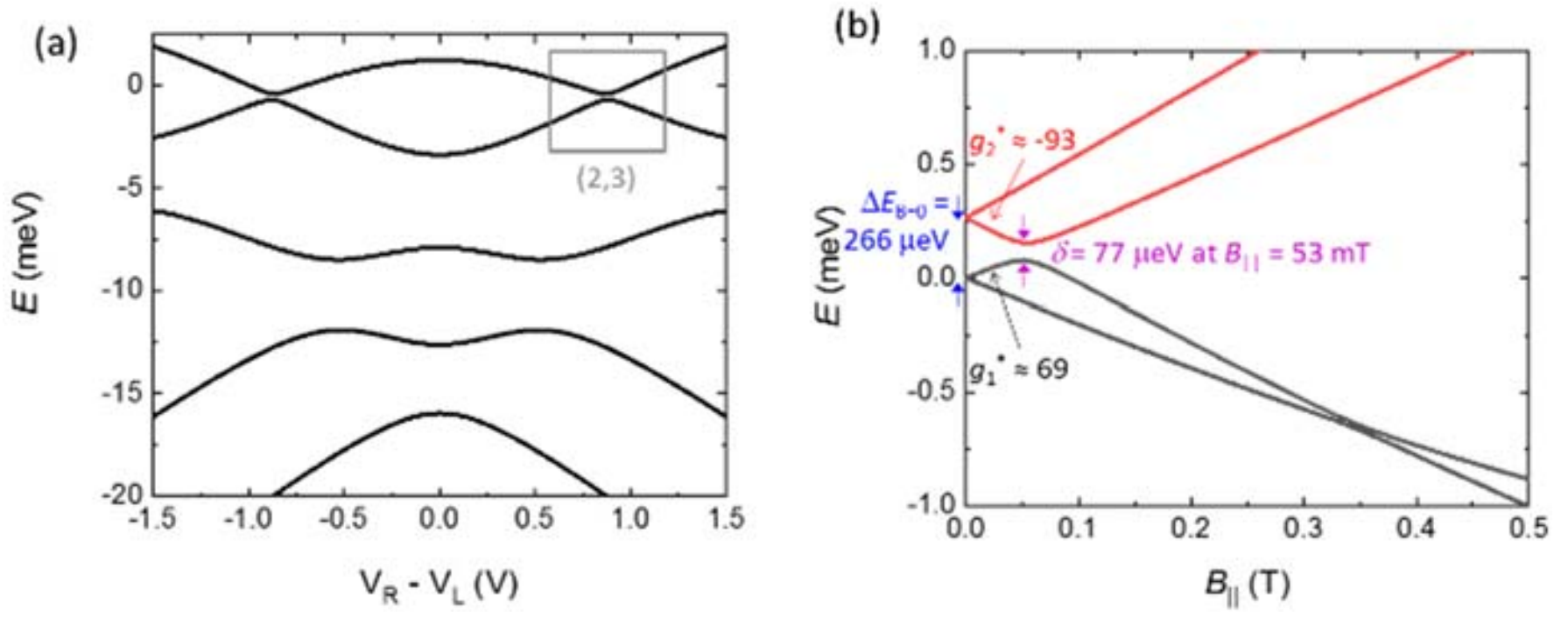

*Figure S2 Numerical simulation of single-electron states. (a) The lowest energy states as functions of the difference of the side gate voltages $V_R - V_L$. Here, B=0 and the spin states are therefore degenerate. The states at the (2,3) crossing (marked by the gray rectangle) are studied in detail for comparison with the experiment. (b) The states evolve with external magnetic field along the nanowire axis $B_{||}$. Fitting parameters are tuned to match the energy gap at $B_{||}$ = 0 ($\Delta E_{B=0}$, indicated by blue arrows), the energy split related to backscattering $\delta$ and the anti-crossing $B_{||}$ field (indicated by purple arrows), and the g-factors $g_1^*$ and $g_2^*$ between the experiment and the simulation.*

| | Experiment | Simulation |
|---|---|---|
| $\delta$ (μeV) | 50 | 77 |
| $\Delta E_{B=0}$ (μeV) | 240 | 266 |
| $g_1^*$ | 59 | 69 |
| $g_2^*$ | -83 | -93 |
| Material parameters | | |
| $\delta\rho$ (C m$^{-2}$) | $0.26 \times 10^{-3}$ | |
| $\rho_{ZB}$ (C m$^{-2}$) | $1.10 \times 10^{-3}$ | |
| $\rho_{WZ}$ (C m$^{-2}$) | $0.24 \times 10^{-3}$ | |
| ZB/WZ conduction band offset (meV) | 120 | |
| $g^*_{spin}$ | -11 | |
| $\alpha$ ( meV nm ) | 16.3 | |

*Table S1 Comparison between the energy splits and effective g-factors extracted from the experiment and the numerical simulation. The shown material parameters are used in the numerical simulation.*

# Tight-binding model of the quantum ring

In this section, we describe an alternative way to model the two half-ring QDs based on a tight-binding chain with periodic boundary conditions, as depicted in Fig. S3(a). The Hamiltonian is:

$$H = H_{ring} + H_{SO} + H_Z, \tag{S6}$$

where $H_{ring}$ contains the kinetic and chemical potential terms and $H_{SO}$ and $H_Z$ are the spin-orbit (SO) interaction and Zeeman energy terms respectively. The explicit forms of the Hamiltonians are:

$$H_{ring} = (2\mathrm{t} - \mu) \sum_{\sigma,j=1}^{N} c_{j,\sigma}^{\dagger} c_{j,\sigma} - t \sum_{\sigma,j=1}^{N} \{\tilde{c}_{j,\sigma}^{\dagger} \tilde{c}_{j+1,\sigma} + H.C.\},$$

$$H_{SO} = t_{so} \sum_{\sigma,\sigma',j=1}^{N} \{-i(\sigma_z)_{\sigma\sigma'} \tilde{c}_{j,\sigma}^{\dagger} \tilde{c}_{j+1,\sigma'} + H.C.\}, \tag{S7}$$

$$H_Z = E_Z \sum_{\sigma,\sigma',j=1}^{N} (\sigma_z)_{\sigma\sigma'} c_{j,\sigma}^{\dagger} c_{j,\sigma'},$$

where $c_{j,\sigma}^{\dagger}$ ($c_{j,\sigma}$) creates (annihilates) an electron with spin $\sigma = \uparrow, \downarrow$ on site $j$, $\mu$ is the chemical potential, $N$ is the total number of sites, $t = \hbar^2/2m^* d^2$ is the kinetic energy associated with hopping between neighboring sites ($m^*$ being the effective electron mass and $d$ the lattice constant), $t_{so} = \alpha/2d$ is the energy associated with the SO interaction ($\alpha$ being the SO strength) and $E_Z = g_{spin}^* \mu_B |\vec{B}|/2$ is the Zeeman energy ($g_{spin}^*$ being the effective $g$-factor without orbital contributions, $\mu_B$ the Bohr magneton and $\vec{B}$ the applied magnetic field). The operators $\tilde{c}$ include the orbital magnetic field effects:

$$\tilde{c}_j = c_j e^{-i\frac{e\Phi}{\hbar}\frac{j}{N}}, \tag{S8}$$

where $e$ is the elementary charge, $\Phi$ is the total flux through the ring and $c_j$ are the operators without flux contributions. The ring is divided into two QDs, left (L) and right (R), by potential barriers. We chose model parameters to obtain a good match to the experimental data and the 3D simulation. We have used $m_{InAs,ZB}^* = 0.026 m_e$, $\alpha = 2.3$ meV · nm and $g_{bulk}^{InAs} = -11$. The results do not depend on the value of the chemical potential. The potential barriers separating the two QDs are 131 meV high and 3.3 nm wide. The diameter of the ring is 35 nm; this is smaller than the actual nanowire's diameter ($\simeq$ 80 nm) and the choice was made based on the observation that the ring-like states formed, have a diameter of around 1/3 – 2/3 of the nanowire's diameter (Fig. 4h in the main article).

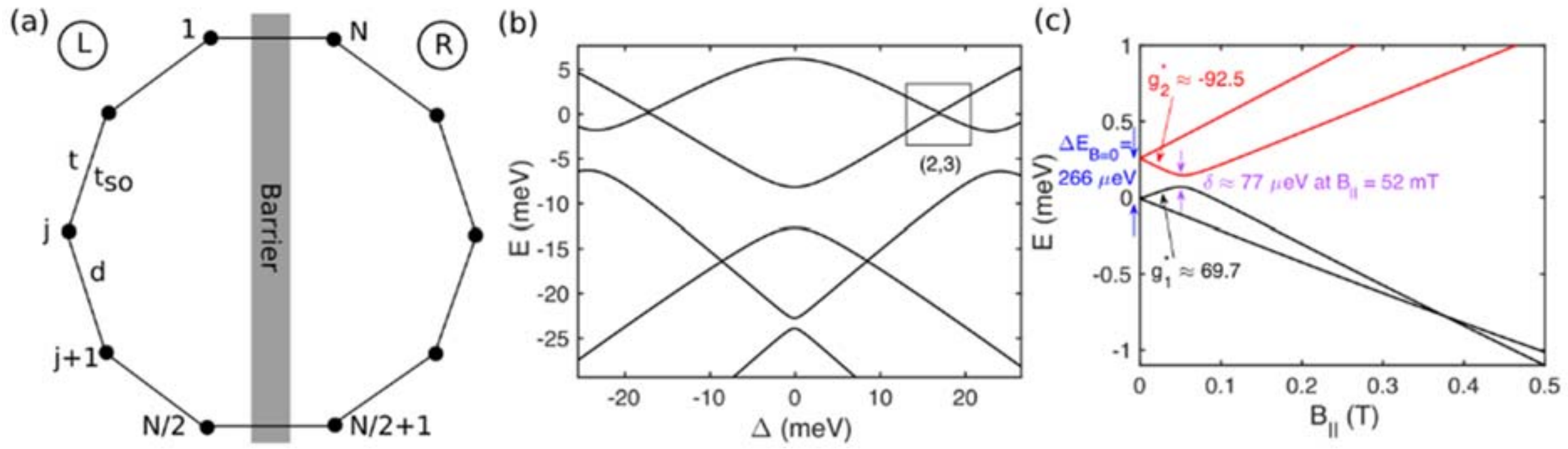

*Figure S3 Tight-binding simulation of single-electron states, cf. Fig. S2. (a) Pictorial representation of the model where the hoppings ($t$, $t_{so}$) and the barrier are depicted. The barrier is implemented considering higher on-site energies for sites between the sites 1, N and N/2, N/2+1 (not shown in the figure) and divides the ring in two half-rings, left (L) and right (R). (b) The lowest energy states as a function of the asymmetry Δ between the L and R QD at $B = 0$ (the states are two-fold degenerate). The (2,3) anti-crossing, which we analyze further in (c), is shown in the black rectangle. (c) Parallel magnetic field dependence of the states involved in the (2,3) anti-crossing. The energy gap at B = 0, the split at $B_{||}$ = 52 mT and the extracted g* factors for the two Kramers pairs are denoted by blue, magenta, black and red arrows respectively and they are in good agreement with the experimental results and the 3-D nanowire simulations for the chosen parameters.*

The system now consists of two half-ring QDs and applying an electric field $\vec{E}$ we can control which QD levels align. In practice, we include the electric field's effects by introducing an asymmetry between the on-site energies of the left and the right QDs. In Fig. S3(b) we plot the first five energy levels of the two QD system as a function of this asymmetry which we call detuning (Δ). Here $B = 0$, and each energy level is 2-fold degenerate. The anti-crossings at $\Delta = 0$ correspond to situations where the same levels from each QD are aligned. For $\Delta \neq 0$, different levels align. We observe that the energy splitting is in general larger for the (odd,odd) and (even,even) crossings than for the (odd,even) and (even,odd) ones. This is crucial to explain the formation of ring-like states and we address this point later in this section. We focus on the (2,3) anti-crossing where the large $g*$ values were observed experimentally, and redefine the "zero" of our detuning at $\Delta \simeq 16.916$ meV. We note that a disorder parameter of $D = 1.365$ meV was also included by adding a random energy between -1.365 and +1.365 meV at each site.

In Fig. S3(c) we plot the evolution of the energies of the states involved in the (2,3) anti-crossing with a magnetic field parallel to the nanowire axis (and thus perpendicular to the considered double QDs system, Fig. S3(a)). The energy splitting at $B = 0$ is $\Delta E_{B=0} \simeq 266\ \mu eV$ while the $g*$ extracted for the first and second Kramers pairs are 69.7 and -92.5 respectively. At $B_{||} = 52$ mT we observe an avoided crossing and the states split by an energy $\delta = 77$ μeV. $\delta$ is induced by the disorder $D$ which quantifies the quality of the ring. Thus, with the chosen parameters, the simple tight-binding chain qualitatively and quantitatively reproduces the experimental data and the 3D nanowire simulation.

The enhancement of $g*$ for the (2,3) anti-crossing is attributed to orbital contributions stemming from the formation of ring-like states. Calculating the expectation value of the angular momentum operator $L = \langle\Psi|\hat{L}|\Psi\rangle$ for the states involved in the (2,3) anti-crossing we find $L \approx 1.075\ \hbar$, a result compatible with the "ring" picture. In contrast, $L \approx 0.01\ \hbar$ for the states involved in the (2,2) anti-crossing, where the orbital contributions are negligible.

We stress here that the extracted $g^*$ values are very sensitive to detuning and they decrease fast away from the detuning corresponding to the (2,3) anti-crossing. This enables the manipulation of the magnetic response of the system by electrical means in a much more efficient way compared to a quantum ring without barriers. To illustrate this point, we plot (Fig. S4) the detuning dependence of $g^*$ for a ring without barriers and for the half-ring QDs. The plotted $g^*$ is normalized by the maximum $g^*$ value ($g^*_{max}$) obtained for the proper detuning value for each system. We notice that for $\Delta = 1$ meV the g* values for the QDs system have dropped to $\approx 0.1\ g^*_{max}$ and $\approx 0.3\ g^*_{max}$ for the lower and upper Kramers pair respectively, whereas $g^*_{ring}$ is practically unaffected and we would have to go to much larger $\Delta$ values to observe a decrease. We conclude that even though large effective $g$-factors can also be extracted for a ring without barriers, the two half-ring QD system is significantly more favorable for electrical manipulation of $g^*$.

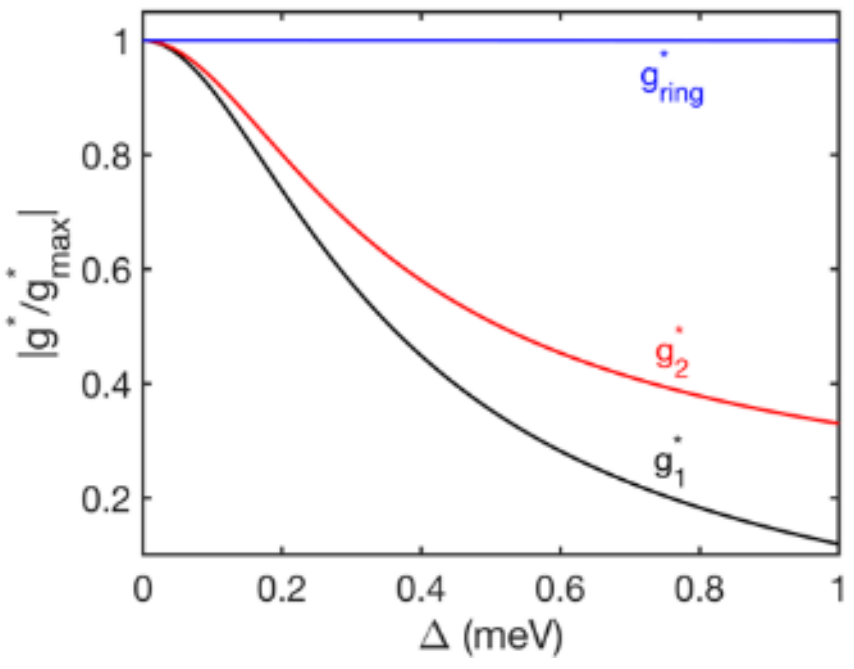

*Figure S4 Detuning dependence of the normalized $g^*$ for a ring without barriers (blue) and for the system of two half-ring QDs (black for lower and red for upper Kramers pair). $g^*_{ring}$ is practically unaffected for the depicted range of detuning, while $g^*_1$ and $g^*_2$ decay fast.*

We now turn to a perturbation theory analysis to gain further insight into why the ring-like states only form for the (even,odd) and (odd,even) crossings. The unperturbed system consists of completely separated L and R QDs, which are one-dimensional quantum wells, their states being described by the usual sinusoidal wave functions. In the TB description the wave functions for the L QD can be written as:

$$|\Psi^{(0)}_{L,n}\rangle = \sum_{\sigma,j=1}^{N/2} a_{L,n,j,\sigma} |\phi_{L,j,\sigma}\rangle, \tag{S9}$$

where $|\phi_{L,j,\sigma}\rangle$, $a_{n,L,j,\sigma}$ are the wavefunction and the wavefunction coefficient at site $j$ and for spin component $\sigma = \uparrow, \downarrow$ and $n$ is the number of the state in the QD. Similarly, for the R QD:

$$|\Psi_{R,n}^{(0)}\rangle = \sum_{\sigma,j=\frac{N}{2}+1}^{N} a_{R,n,j,\sigma}|\phi_{R,j,\sigma}\rangle. \tag{S10}$$

We note that the standard particle in a box wave functions obey:

$$a_{L,n,1,\sigma} = a_{L,n,N/2,\sigma} = a \qquad for\ n\ odd,$$

$$a_{L,n,1,\sigma} = -a_{L,n,N/2,\sigma} = a \qquad for\ n\ even,$$

$$a_{R,n,\frac{N}{2}+1,\sigma} = a_{R,n,N,\sigma} = a \qquad for\ n\ odd,$$

$$a_{R,n,\frac{N}{2}+1,\sigma} = a_{R,n,N,\sigma} = a \qquad for\ n\ even, \tag{S11}$$

where $a$ can be chosen real and positive.

When two levels in the double QDs system are aligned, they form a degenerate subspace. We now consider a perturbation in the form of a coupling between the sites *1,N* and *N/2,N/2+1*

$$\delta H = -t\left\{c_{\frac{N}{2},\uparrow}^{\dagger}c_{\frac{N}{2}+1,\uparrow} + c_{\frac{N}{2},\downarrow}^{\dagger}c_{\frac{N}{2}+1,\downarrow} + c_{N,\uparrow}^{\dagger}c_{1,\uparrow} + c_{N,\downarrow}^{\dagger}c_{1,\downarrow} + H.C.\right\}$$
$$+t_{so}\left\{-ic_{\frac{N}{2},\uparrow}^{\dagger}c_{\frac{N}{2}+1,\uparrow} + ic_{\frac{N}{2},\downarrow}^{\dagger}c_{\frac{N}{2}+1,\downarrow} - ic_{N,\uparrow}^{\dagger}c_{1,\uparrow} + ic_{N,\downarrow}^{\dagger}c_{1,\downarrow} + H.C.\right\}, \tag{S10}$$

according to (S7). At $B$=0 the spin states are degenerate and in the following we focus on one Kramers pair, choosing to work with spin up, as the spin degeneracy cannot be broken with the considered perturbation.

For the (2,2) anti-crossing the aligned states are $|\Psi_{L,2}^{(0)}\rangle$ and $|\Psi_{R,2}^{(0)}\rangle$, thus $E_{L,2}^{0} = E_{R,2}^{0} = E_{(2,2)}^{(0)}$. For the matrix elements of $\delta H$ in the degenerate subspace we find:

$$\delta H_{11} = \left\langle \Psi_{L,2}^{(0)} \right| \delta H \left| \Psi_{L,2}^{(0)} \right\rangle = 0$$

$$\delta H_{12} = \left\langle \Psi_{L,2}^{(0)} \right| \delta H \left| \Psi_{R,2}^{(0)} \right\rangle = -a^2(-t) - a^2(-t) = 2a^2 t$$

$$\delta H_{21} = \left\langle \Psi_{R,2}^{(0)} \right| \delta H \left| \Psi_{L,2}^{(0)} \right\rangle = -a^2(-t) - a^2(-t) = 2a^2 t$$

$$\delta H_{22} = \left\langle \Psi_{R,2}^{(0)} \right| \delta H \left| \Psi_{R,2}^{(0)} \right\rangle = 0. \tag{S11}$$

Diagonalizing $\delta H$ we obtain the first order energy corrections and the corresponding linear combinations of the unperturbed wavefunctions:

$$\left|\Psi^{(0)}_{\alpha,(2,2)}\right\rangle = \frac{1}{\sqrt{2}}\left(\left|\Psi^{(0)}_{L,2}\right\rangle - |\Psi^{(0)}_{R,2}\rangle\right), \qquad E^{(1)}_{\alpha,(2,2)} = E^{(0)}_{(2,2)} - 2a^2 t,$$

$$\left|\Psi^{(0)}_{\beta,(2,2)}\right\rangle = \frac{1}{\sqrt{2}}\left(\left|\Psi^{(0)}_{L,2}\right\rangle + |\Psi^{(0)}_{R,2}\rangle\right), \qquad E^{(1)}_{\beta,(2,2)} = E^{(0)}_{(2,2)} + 2a^2 t. \tag{S12}$$

We follow a similar procedure for the (2,3) anti-crossing. Now $E^0_{L,2} = E^0_{R,3} = E^{(0)}_{(2,3)}$, since the aligned states are $|\Psi^{(0)}_{L,2}\rangle$ and $|\Psi^{(0)}_{R,3}\rangle$. The first order energy corrections and the corresponding linear combination of the unperturbed wavefunctions are:

$$\left|\Psi^{(0)}_{\alpha,(2,3)}\right\rangle = \frac{1}{\sqrt{2}}\left(\left|\Psi^{(0)}_{L,2}\right\rangle - i|\Psi^{(0)}_{R,3}\rangle\right), \qquad E^{(1)}_{\alpha,(2,3)} = E^{(0)}_{(2,3)} + 2a^2 t_{so},$$

$$\left|\Psi^{(0)}_{\beta,(2,3)}\right\rangle = \frac{1}{\sqrt{2}}\left(\left|\Psi^{(0)}_{L,2}\right\rangle + i|\Psi^{(0)}_{R,3}\rangle\right), \qquad E^{(1)}_{\beta,(2,3)} = E^{(0)}_{(2,3)} - 2a^2 t_{so}. \tag{S14}$$

The above results can explain the large difference between the (2,2) and (2,3) energy splittings visible in Fig. S3(b). In the (2,2) case the splitting is $4a^2t$ which is much larger than $4a^2t_{so}$ in the (2,3) case. The (even,odd) splittings are thus induced by $t_{so}$ and vanish for $t_{so} = 0$. From the TB results we can extract $4a^2t \doteq 4\tilde{t} = 4.528$ meV and $4a^2t_{so} \doteq 4\tilde{t}_{so} = 0.266\, meV$. It is now straightforward to write down the first order corrections to the wave functions (up to a normalization factor):

$$\left|\Psi^{(1)}_{\alpha(\beta),(2,2)}\right\rangle = \left|\Psi^{(0)}_{\alpha(\beta),(2,2)}\right\rangle \pm \sqrt{2}\left\{\tilde{t}\sum_p \frac{1}{E^{(0)}_{p_{LE}} - E^{(0)}_{(2,2)}}|p_{LE}{}^{(0)}\rangle - i\tilde{t}_{so}\sum_p \frac{1}{E^{(0)}_{p_{LO}} - E^{(0)}_{(2,2)}}|p_{LO}{}^{(0)}\rangle\right\}$$

$$-\sqrt{2}\left\{\tilde{t}\sum_p \frac{1}{E^{(0)}_{p_{RE}} - E^{(0)}_{(2,2)}}|p_{RE}{}^{(0)}\rangle - i\tilde{t}_{so}\sum_p \frac{1}{E^{(0)}_{p_{RO}} - E^{(0)}_{(2,2)}}|p_{RO}{}^{(0)}\rangle\right\}, \tag{S14}$$

$$\left|\Psi^{(1)}_{\alpha(\beta),(2,3)}\right\rangle = \left|\Psi^{(0)}_{\alpha(\beta),(2,3)}\right\rangle \mp \sqrt{2}\left\{\tilde{t}_{so}\sum_p \frac{1}{E^{(0)}_{p_{LE}} - E^{(0)}_{(2,3)}}|p_{LE}{}^{(0)}\rangle - i\tilde{t}\sum_p \frac{1}{E^{(0)}_{p_{LO}} - E^{(0)}_{(2,3)}}|p_{LO}{}^{(0)}\rangle\right\}$$

$$-\sqrt{2}\left\{\tilde{t}\sum_p \frac{1}{E^{(0)}_{p_{RE}} - E^{(0)}_{(2,3)}}|p_{RE}{}^{(0)}\rangle - i\tilde{t}_{so}\sum_p \frac{1}{E^{(0)}_{p_{RO}} - E^{(0)}_{(2,3)}}|p_{RO}{}^{(0)}\rangle\right\}$$

$$\mp\frac{1}{2}\left\{\frac{\tilde{t}^2}{\tilde{t}_{so}}\sum_p \frac{1}{E^{(0)}_{p_{RE}} - E^{(0)}_{(2,3)}} + \tilde{t}_{so}\sum_p \frac{1}{E^{(0)}_{p_{RO}} - E^{(0)}_{(2,3)}}\right\}\left|\Psi^{(0)}_{\beta(\alpha),(2,3)}\right\rangle$$

$$\mp\frac{1}{2}\left\{-\tilde{t}_{so}\sum_p \frac{1}{E^{(0)}_{p_{LE}} - E^{(0)}_{(2,3)}} - \frac{\tilde{t}^2}{\tilde{t}_{so}}\sum_p \frac{1}{E^{(0)}_{p_{LO}} - E^{(0)}_{(2,3)}}\right\}\left|\Psi^{(0)}_{\beta(\alpha),(2,3)}\right\rangle, \tag{S15}$$

where index $p_{L(R)E(O)}$ refers the $p^{th}$even (odd) state outside the degenerate subspace in QD L (R). Examining the above expressions we notice that in the (2,2) case even and odd wave functions are added multiplied with $\tilde{t}$ and $i\tilde{t}_{so}$ respectively. Adding odd purely imaginary wavefunctions to the even $\left|\Psi^{(0)}_{\alpha(\beta),(2,2)}\right\rangle$ contributes in making the zeros in the magnitude of $\left|\Psi^{(1)}_{\alpha(\beta),(2,2)}\right\rangle$ finite, but since $\tilde{t}_{so} \ll \tilde{t}$ the effect is not noticeable. The situation is different in the (2,3) case. Odd imaginary wave functions are added to $\left|\Psi^{(0)}_{L,2}\right\rangle$ and even real wavefunctions are added to $i|\Psi^{(0)}_{R,3}\rangle$, in both cases multiplied with $\tilde{t}$. This has the effect that the magnitude of $\left|\Psi^{(1)}_{\alpha(\beta),(2,3)}\right\rangle$ varies less in real space. In this sense the wave functions in the (2,3) case are more like the eigenstates of a perfect ring, which have a constant magnitude.

# Additional information about sample A

Figures S5a-b show overview measurements of sample A at $V_{BG} = 1$ V, and $V_{BG} = -1$ V, respectively. The orbital crossings which are presented in this supporting material are highlighted. As in the main article, we label the crossings ($O_L$,$O_R$), where $O_L$ and $O_R$ is the orbital number of the left and right QD. Crossings (2,3), (1,1) and (2,2) were investigated at $V_{BG} = 1$ V, while crossing (4,1) and crossing (5,2) were studied at $V_{BG} = 0$ V, and -1.5 V, respectively. Since changing $V_{BG}$ affects the tunnel couplings between the left and right QD, we note that some crossings are only clearly visible for specific gate voltage ranges.

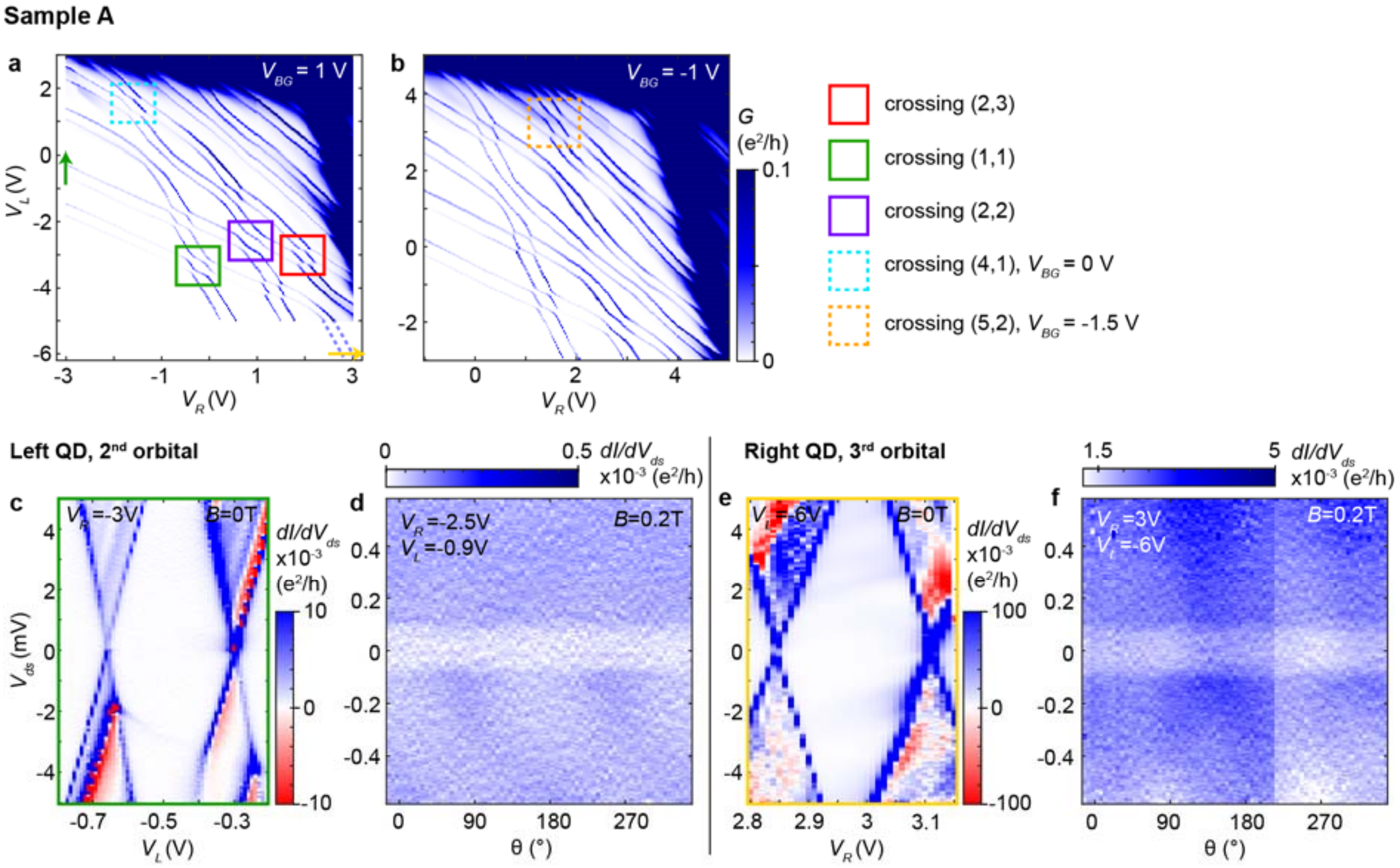

*Figure S5* ***Overview measurements and behaviour of single orbitals. a-b*** *Conductance measurements over large sidegate voltage ranges for* $V_{BG} = 1$ *V, and -1 V, respectively. The crossings which are discussed in this supporting material are highlighted.* ***c*** *Measurement of* $dI/dV_{ds}$ *versus* $V_{ds}$ *as a function of* $V_L$ *for the 2nd orbital of the left QD (green vector).* ***d*** *Measurement of* $dI/dV_{ds}$ *versus* $V_{ds}$ *as a function of B-field direction.* ***e-f*** *Corresponding measurements for the 3rd orbital of the right QD.*

The behaviour of two isolated orbitals is presented in Figs. S5c-f. We focus on the 2nd orbital from the left QD, and the 3rd orbital from the right QD, which are the relevant orbitals for crossing (2,3). Figure S5c shows transport as a function of $V_L$ for $O_L = 2$, when there are zero electrons in the right QD. The charging energy $E_c$ and lever arm $\alpha_{VL/L}$ of $V_L$ on the left QD can be extracted from the height and the width of the Coulomb diamond. The lever arm $\alpha_{VR/L}$ of $V_R$ on the left QD can then be calculated using the slope of the conductance lines Fig. S5a. Similar analysis can be done for the right QD using Fig. S5e, and all parameters for both QDs are shown in Table S2.

| | $E_c$ (meV) | $\alpha_{VL/L}$ (meV/V) | $\alpha_{VR/L}$ (meV/V) | $\alpha_{VL/R}$ (meV/V) | $\alpha_{VR/R}$ (meV/V) |
|---|---|---|---|---|---|
| Left QD | 9.9 | 28 | 19 | - | - |
| Right QD | 7.6 | - | - | 11 | 29 |

*Table S2 Charging energies and leverarms of the QDs of sample A.*

Knowing the lever arms, an orbital separation > 30 meV for both QDs can be calculated from Fig. S5a. Finally, we note that the Zeeman splitting of the single QD orbitals show almost no rotation dependence (Figs. S5d,f), and a $g^*$ of ~ 8 can be calculated for both orbitals.

## Crossing (2,3) – additional information

The main article shows the most important transport characteristics of crossing (2,3). Here we present additional data from the same crossing.

**1e regime**

We start by showing data from the 1e regime. To facilitate the discussion, we present the numerical calculation of the state energies as function of $B_{||}$ again in Fig. S6a (same as Fig. 2b in the main article). Figures S6b-c show transport as a function of $B$-field orientation for $B$ = 0.05 T, and $B$ = 0.5 T ($B$ = 0.2 T is presented in the main article). While a large anisotropy of $g^*$ can be observed for any $B$ > 0, it is interesting to note the change in bowing of the smallest energy gap when going to higher $B$-fields. This is a consequence of a change in orbital momentum sign of the first excited state when the magnetic flux that penetrates the ring is sufficiently large. Figures S6d-h show transport as a function of detuning in the 1e regime for different $B$-field strength. This series visualizes the hybridization of the 1$^{st}$ and the 2$^{nd}$ excited states (ES1, ES2) at zero detuning. For $B_{||}$ < 55 mT, the gap between ES1 and the ground state (GS) increases with the $B$-field, and allows to extract $g_1^*$ directly (which was done in the main article). For $B_{||}$ > 55 mT the gap at zero detuning decreases with increasing $B$-field, since the two lowest states now have the same orbital momentum but the spin of excited state is favorable at higher fields. Eventually, for $B_{||}$ > 0.35 T, the ground state spin of the ring changes and is the same as for the single QD orbitals. Therefore no change in ground state is observed when detuning the orbitals at $B$ = 0.5 T (Fig. S6h).

Using the lever arms we can now also convert $\Delta V_{L,R}$ to an energy, which can then be compared with $\Delta_{orb}$. In Figures S6d-f, $V_L$ is changed from -3.15 V to -3.32 V, and simultaneously the right sidegate is changed from 1.74 V to 1.9 V.

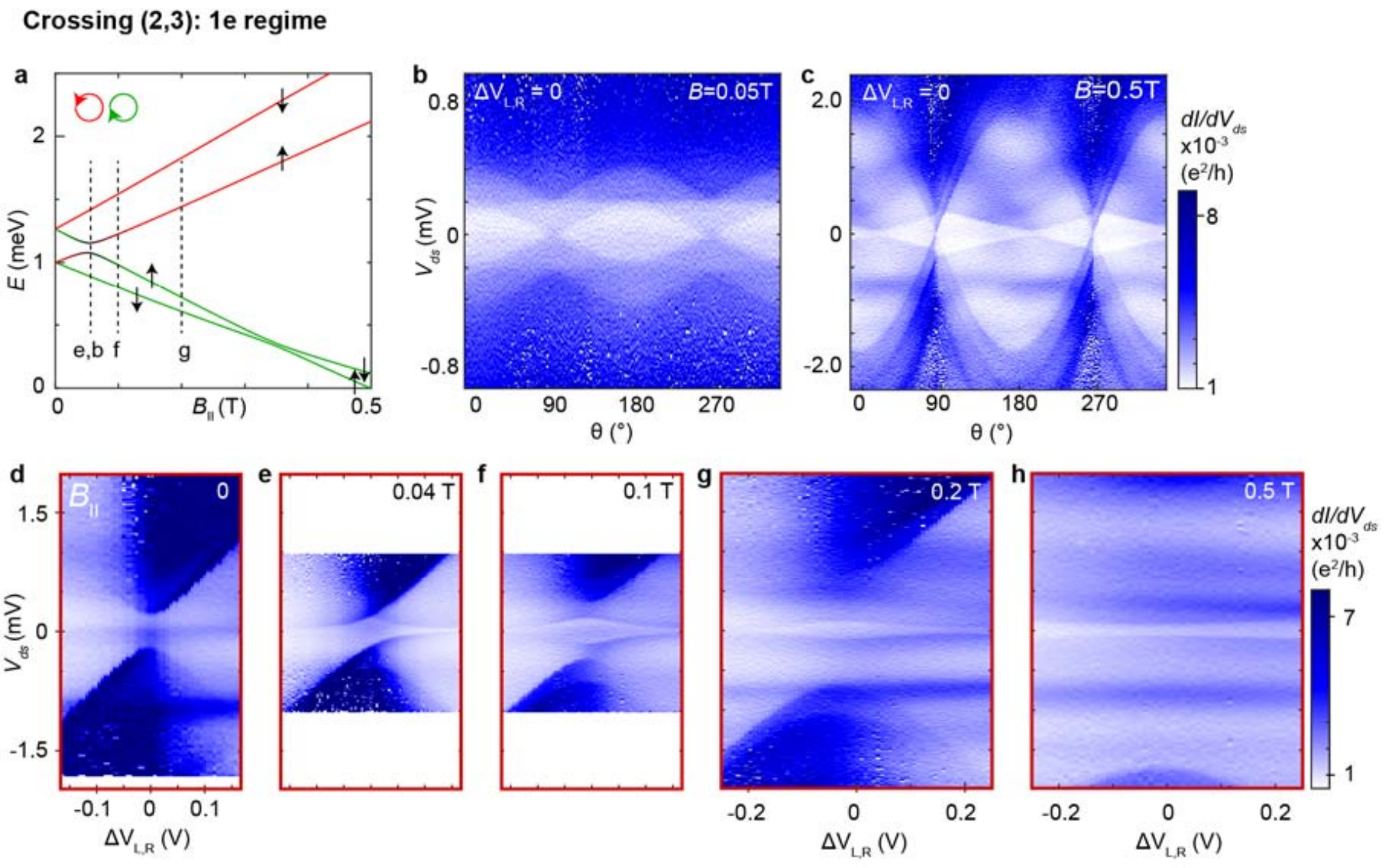

*Figure S6* ***Additional transport data from the 1e regime of crossing (2,3). a*** *Numerical calculation of state energies as a function of $B_{||}$ (same as Fig. 2b in the main article).* ***b-c*** *Measurement of $dI/dV_{ds}$ versus $V_{ds}$ as a function of B-field direction at zero detuning, for B = 0.05T, and 0.5T.* ***d-h*** *Measurement of $dI/dV_{ds}$ versus $V_{ds}$ recorded along the red detuning vector in the 1e regime, for different $B_{||}$ -field strengths.*

### 3e regime

As explained in the main article, the 3e regime shows a different behavior compared to the 1e regime. In Fig. S7 we present additional data supporting this observation. Transport as a function of $B_{\perp}$ (Fig. S7a) shows a very weak $B$-field dependence, similar to what has been shown for the 1e regime. A corresponding suppression of $g^*$ at zero detuning can be observed in Fig. S7b. However, the magnetic field rotation for $B$ = 0.2 T (Fig. S7b) shows a clearly different behavior, in particular there is no opposite bowing for the ES1-GS gap. This can also be observed in the $B$-field series of the detuning measurements (Figs. S7d-g): the gap at zero detuning continuously increases with $B$-field strength, and is always larger compared to the gap of the single orbitals. This also means that a change of the spin ground state (Figure 4d of the main article) does not occur in the 3e regime, neither as function of $B_{||}$ nor detuning.

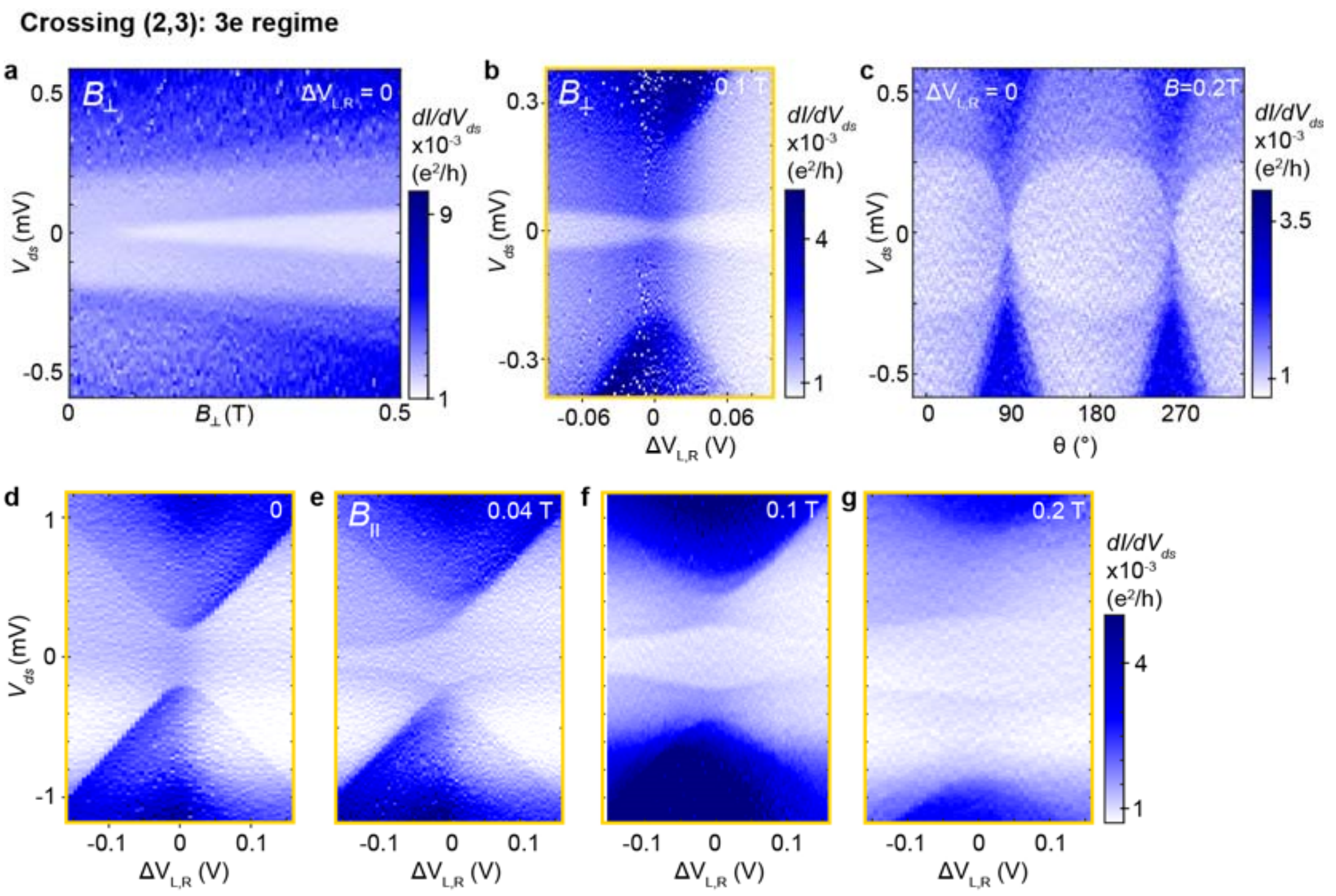

*Figure S7* ***Additional transport data from the 3e regime of crossing (2,3).*** ***a-c*** *Measurement of $dI/dV_{ds}$ versus $V_{ds}$ as a function of $B_\perp$ at zero detuning.* ***d-i*** *Measurement of $dI/dV_{ds}$ versus $V_{ds}$ recorded along the yellow detuning vector in the 3e regime, for $B_\perp = 0.1$ T.* ***c*** *Measurement of $dI/dV_{ds}$ versus $V_{ds}$ as a function of B-field direction at zero detuning for $B = 0.2$ T.* ***d-g*** *Measurement of $dI/dV_{ds}$ versus $V_{ds}$ recorded along the yellow detuning vector in the 1e regime, for different $B_{||}$ -field strengths.*

## B-field dependence of the overview

Finally, we present how the ground states of crossing (2,3) depend on the *B*-field strength and orientation (Fig. S8). As highlighted in the main article, one characteristic of a crossing where the hybridization leads to the formation of ring-like states, is that the corners of the crossing appear very sharp. The strong suppression of the hybridization gap is suggested to be a consequence of tunnel-coupling an even and an odd QD orbital in two points. Another characteristic is that the energy of the ground states dramatically changes when increasing $B_{||}$ from 0 to 0.5 T (Figs. S8a-d). This stands in sharp contrast to Fig. S8e, which shows that $B_\perp = 0.5$ T effectively leaves the states unchanged compared to $B = 0$. We would like to highlight that this observation provides an easy way to screen for the occurrence of ring-like states in orbital crossings: ring-like states can be identified by comparing the honeycomb pattern of the crossing at both parallel and perpendicular magnetic field.

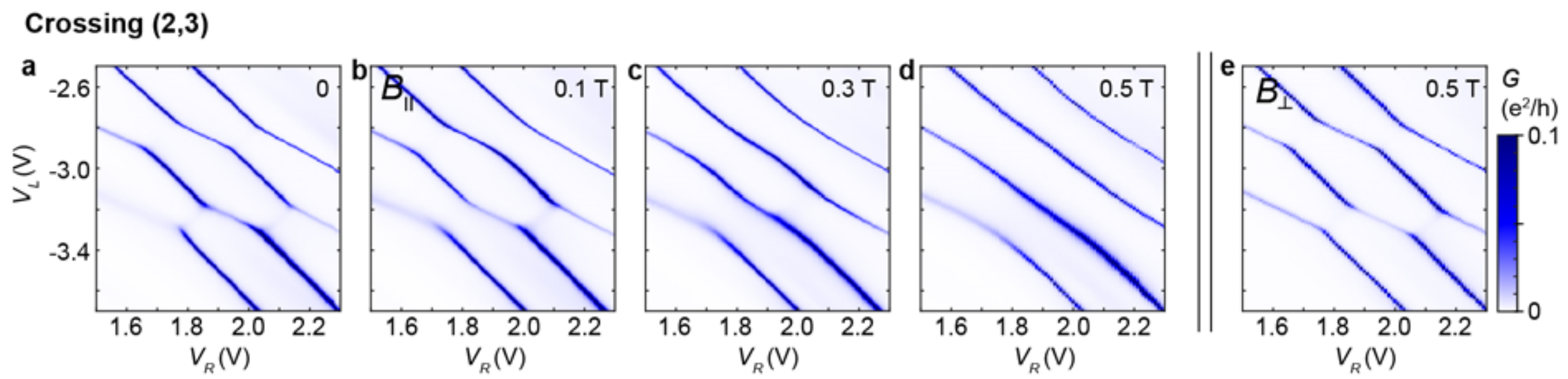

*Figure S8* ***B-field dependence of crossing (2,3). a-d*** *Conductance as a function of sidegate voltages for* $B_{||}$ *= 0, 0.1 T, 0.3 T, and 0.5T, respectively.* ***e*** *Corresponding measurement for* $B_{\perp}$ *= 0.5 T.*

## Ring states in other even/odd or odd/even crossings

Figure S9 shows transport data from another even/odd and an odd/even crossing. In the overview measurements (Figs. S9a,f) very sharp corners can be observed, similar to crossing (2,3), which is related to the strongly suppressed hybridization gap. We study the 3e regime of crossing (4,1), and the 1e regime of crossing (5,2) in more detail. For both crossings, transport at zero detuning as a function of $B_{||}$ (Figs. S9b,g) shows two excited states that rapidly increase in energy, and a huge anisotropy upon magnetic field rotation can be observed (Figs. S9c,h), corresponding to a large orbital contribution to the effective *g*-factor (g*~50-75 for crossing (4,1), and g*~35-43 for crossing (5,2)). Figures S9d-e and S9i-j show transport as a function of detuning for different *B*-field. ES2 and ES3 quickly increase in energy with increasing *B*-field at zero detuning, as expected for ring states. For crossing (4,1) the detuning dependence of the GS-ES1 gap is very small (Fig. S9e), which implies that the hybridization of ES1 and ES2 occurs at very small $B_{||}$-fields due to a small $\Delta_{SOI}$. The GS-ES1 gap for the 1e regime of crossing (5,2) (Fig S9j) shows a similar behavior compared to the 3e regime of crossing (2,3): the gap is the largest at zero detuning, and no ground state change occurs when increasing the detuning. This implies an opposite spin-filling sequence compared to crossing (2,3). However, the small splitting of the Kramers pairs makes it difficult to resolve all the states in the $B_{||}$ sweep, and we therefore discuss this effect in more detail for Sample B.

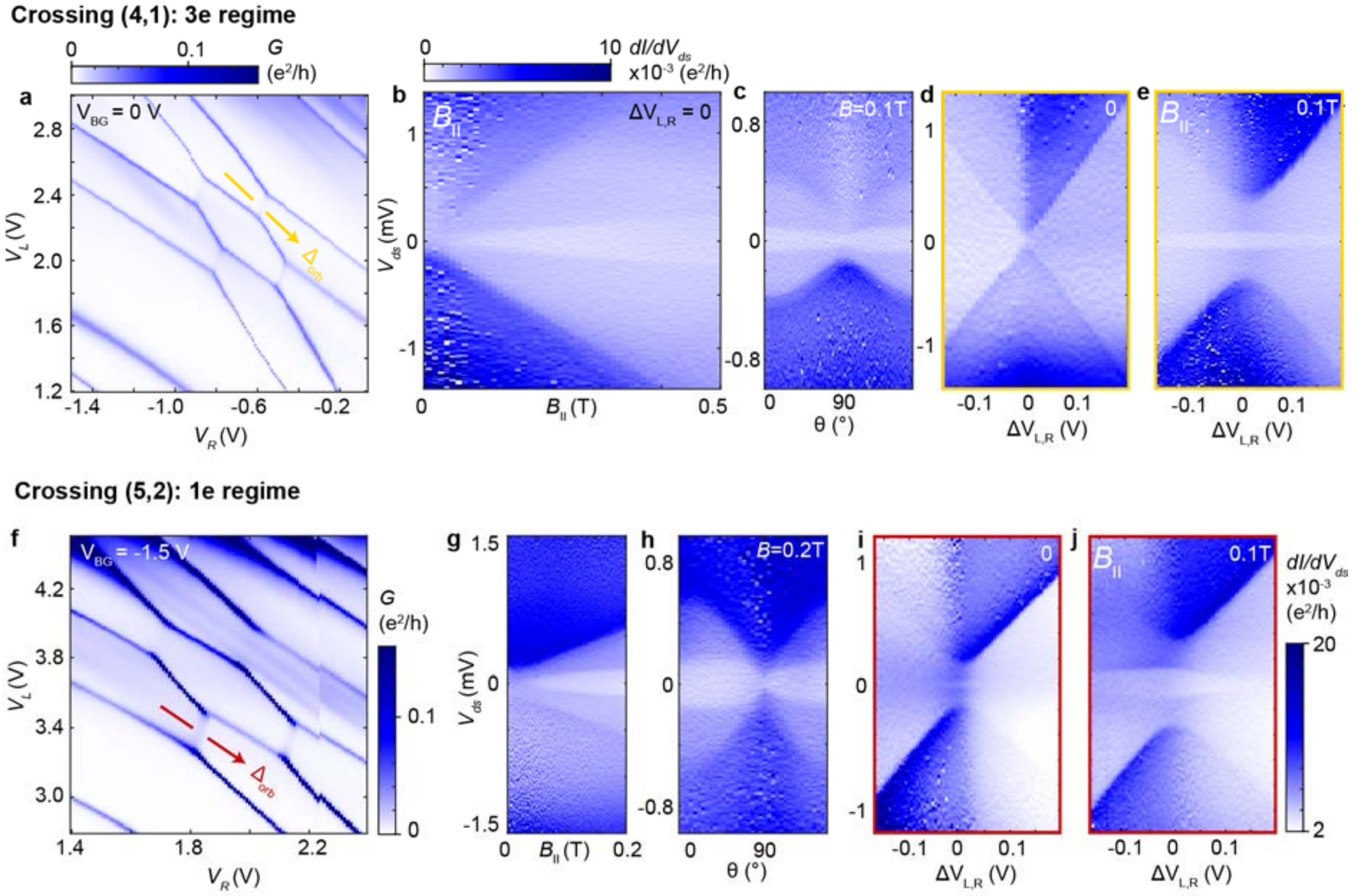

*Figure S9* ***Transport in crossings (4,1) and (5,2). a*** *Conductance of crossing (4,1) as a function of sidegate voltages.* ***b-c*** *Measurement of dI/dV*$_{ds}$ *versus V*$_{ds}$ *at zero detuning in the 3e regime as a function of B*$_{||}$ *and as a function of B-field direction.* ***d-e*** *Measurement of dI/dV*$_{ds}$ *versus V*$_{ds}$ *recorded along the red detuning vector in the 1e regime, for B=0, and B*$_{||}$ *= 0.1 T, respectively.* ***f-j*** *Corresponding measurements for crossing (5,2) in the 1e regime.*

## Absence of ring states in even/even and odd/odd crossings

Our theoretical models predict that perfect rings can only form in the case where and even and an odd orbital of the two QDs are involved. In the case of even/even and odd/odd crossings, the overlap integrals at the barriers have the same sign, leading to a significant hybridization gap. In Fig. S10 we present transport data from crossings (1,1) and (2,2) to support this statement. For both crossings, the overview conductance measurements (Figs. S10a,d) look comparable to measurements of strongly tunnel-coupled parallel QDs (Nilsson2017), which stands in contrast to what has been observed for the crossings discussed in this work so far. In Figs. S10b,e we present transport in the 1e regime as a function of *B*-field direction at zero detuning, and observe that there is almost no rotation dependence of the GS-ES transitions, which is in agreement with no orbital contribution to $g^*$. Accordingly, no change of the gap between GS and ES1 is found when detuning the orbitals along the red gate vector for $B_{||}$ = 0.2 T (Figs. S10c,f).

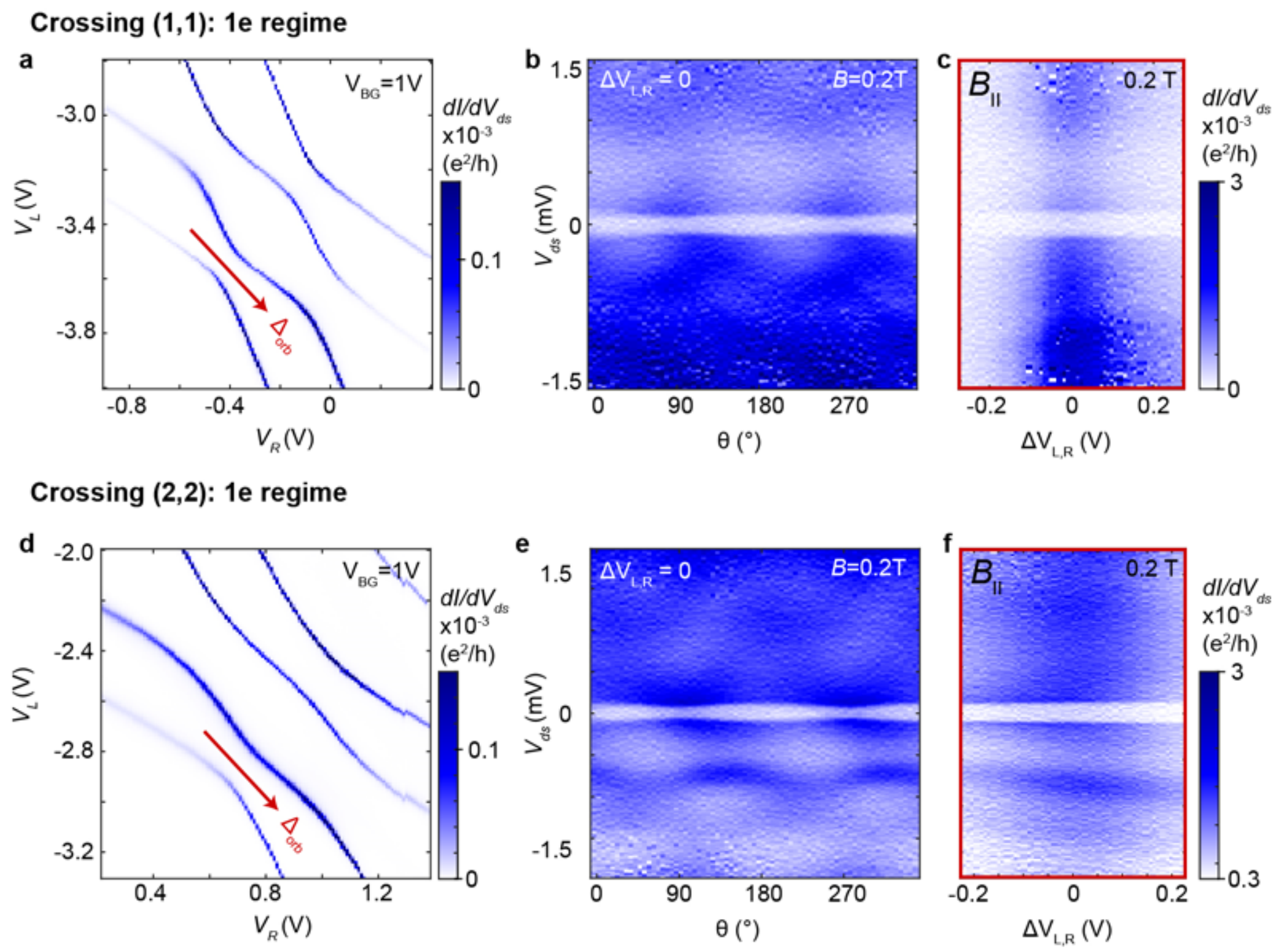

*Figure S10* ***Transport in crossings (1,1) and (2,2). a*** *Conductance of crossing (1,1) as a function of sidegate voltages.* ***b*** *Measurement of dI/dV$_{ds}$ versus V$_{ds}$ at zero detuning in the 1e regime as a function of B-field direction.* ***c*** *Measurement of dI/dV$_{ds}$ versus V$_{ds}$ recorded along the red detuning vector in the 1e regime for $B_{||}$ = 0.2 T.* ***d-f*** *Corresponding measurements for crossing (2,2).*

# Sample B

The emergence of ring states could be reproduced in a second sample (Sample B), with design similar to Sample A. An overview conductance measurement of Sample B is presented in Fig. S11a, and the relevant crossing is shown in Fig. S11b. Transport along the green gate vector at $B_{||}$ = 0.1 T (Fig. S11c) shows weakly outlined Coulomb diamonds, and inelastic co-tunneling transport in the 1e, 2e, and 3e regimes. The gap energy in the 1e regime corresponds to a $g^* \approx 43$, which is much larger compared to the effective *g*-factor of bulk InAs, similar to what has been shown for sample A. In Figs. S11d-g we study the 1e regime in more detail. Transport at zero detuning as a function of $B_{||}$ (Fig. S11d) shows a strong increase of the ES1-GS energy gap with increasing *B*-field, and an anti-crossing with ES2 can be observed at $B_{||} \approx 0.15$T. A huge anisotropy upon magnetic field rotation can be observed in Fig. S11e, corresponding to a large orbital contribution to $g^*$.

Detuning the orbitals along the red gate vector (Figs. S11f-g) shows a strong increase in the gap between GS and ES1 upon formation of the ring states at zero detuning. The gap energy at zero detuning increases continuously with increasing *B*-field, and no ground state spin change can be observed when detuning the orbitals. This behavior is similar to what has been observed for the 3e regime of sample A. To explain this finding, we would like to recall that the spin filling sequence of the ring-like states depends on the sign of the spin-orbit interaction. Crossing (2,3) of Sample A is an example of a crossing for which the ground state at $B$ = 0 and zero detuning is spin-up, which is the un-favored spin-direction for high magnetic fields. This leads to a ground state spin-change with increasing $B_{||}$ in the 1e regime, as well as a ground state spin change when detuning the orbitals at constant magnetic field. The 3e regime, on the other hand, does not show these features. The experimental results of Sample B can therefore be explained by an opposite spin-filling sequence, which is related to an opposite sign of the spin-orbit interaction.

In conclusion, Sample B shows the possibility to create ring-like states, very similar compared to Sample A. In addition, Sample B clearly shows an opposite spin-filling sequence compared to crossing (2,3) of Sample A. However, we would like to note that the spin-filling sequence seems not sample specific but rather depends on the properties of each individual crossing, as a similar behavior was already observed for crossing (5,2) of Sample A.

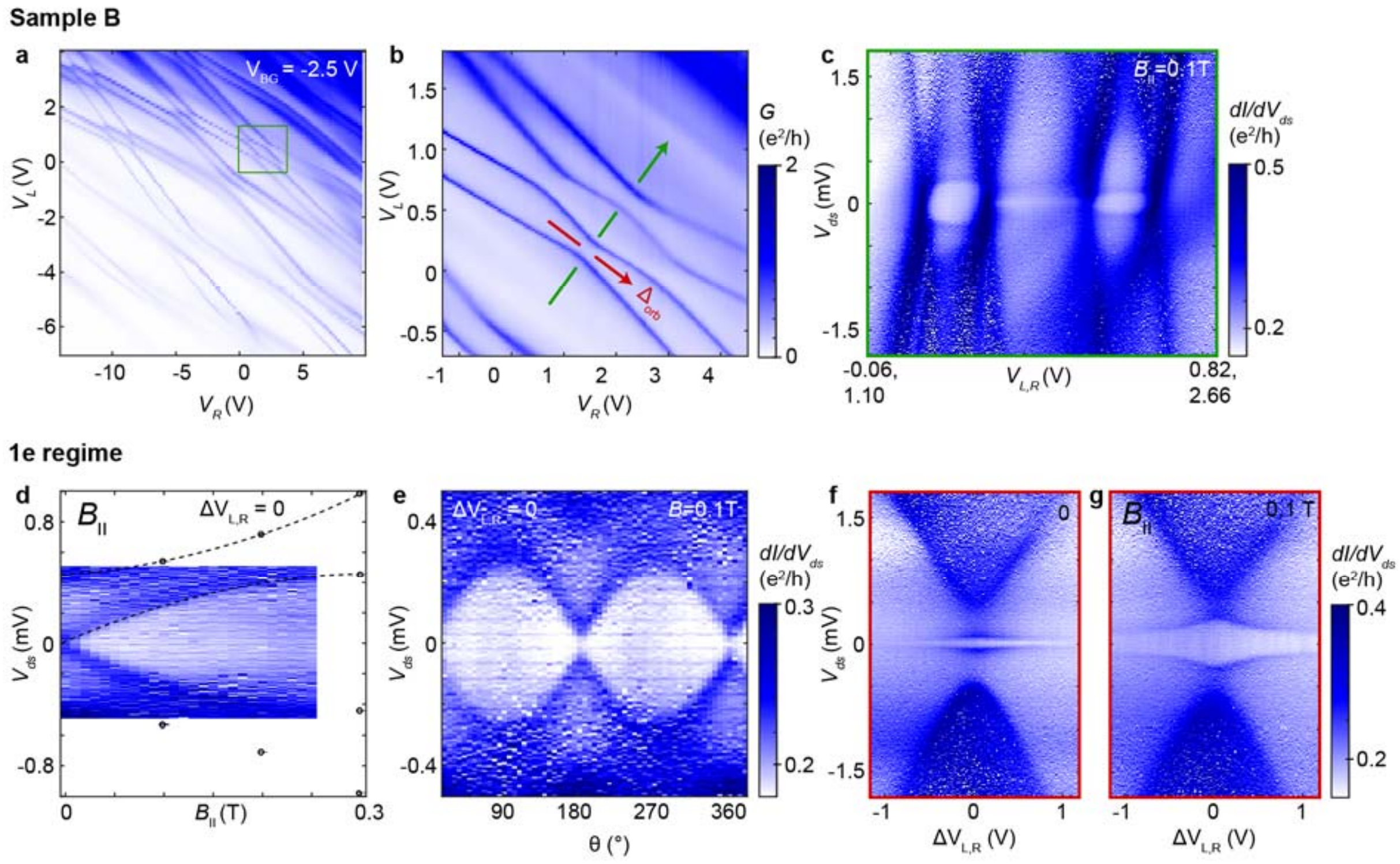

*Figure S11* ***Transport of sample B. a*** *Conductance of Sample B as a function of sidegate voltages.* ***b*** *Overview measurement of the crossing which is investigated in more detail.* ***c*** *Measurement of $dI/dV_{ds}$ versus $V_{ds}$ along the green gate vector.* ***d-e*** *Measurement of $dI/dV_{ds}$ versus $V_{ds}$ at zero detuning in the 1e regime as function of $B_{||}$ and as a function of B-field direction. The datapoints added to* ***d*** *represent gap energies extracted from detuning measurements at different $B_{||}$-field.* ***f-g*** *Measurement of $dI/dV_{ds}$ versus $V_{ds}$ recorded along the red detuning vector in the 1e regime, for $B$ = 0, and $B_{||}$ = 0.1 T, respectively. The finite gap around zero bias in the case of $B$ = 0 can be explained by the ferromagnetic contacts of the device.*